\documentclass{article}

\usepackage[ansinew]{inputenc}  
\usepackage{footnpag}  
\usepackage[ngerman,english]{babel}

\usepackage{graphicx}
\usepackage{amsmath,amssymb}
\usepackage[super,comma,sort&compress]{natbib}

\newcommand{\url}{}
\makeatletter
\newcommand*{\citenumns}[2][]{%
 \begingroup
 \let\NAT@mbox=\mbox
  \let\@cite\NAT@citenum
  \let\NAT@space\NAT@spacechar
  \let\NAT@super@kern\relax
  \renewcommand\NAT@open{[}%
  \renewcommand\NAT@close{]}%
  \cite[#1]{#2}%
  \endgroup
}
\makeatother

\begin{document}

\title{Marcel Grossmann and his contribution to the general theory of relativity}

\author{Tilman Sauer\\[0.5cm]
\footnotesize{Institute for Theoretical Physics}\\[-0.1cm]
\footnotesize{Albert Einstein Center for Fundamental Physics}\\[-0.1cm]
\footnotesize{University of Bern, CH-3012 Bern, Switzerland}\\[-0.1cm]
\footnotesize{and}\\[-0.1cm]
\footnotesize{Einstein Papers Project, California Institute of Technology MC20-7}\\[-0.1cm]
\footnotesize{1200 E California Blvd, Pasadena, CA 91125, USA}\\[-0.1cm]
\footnotesize{E-mail: tilman@caltech.edu}}

\date{Version of \today}

\maketitle

\begin{abstract}
This article reviews the biography of the Swiss mathematician Marcel Grossmann (1878--1936) and his contributions to the emergence of the general theory of relativity. The first part is his biography, while the second part reviews his collaboration with Einstein in Zurich which resulted in the Einstein-Grossmann theory of 1913. This theory is a precursor version of the final theory of general relativity with all the ingredients of that theory except for the correct gravitational field equations. Their collaboration is analyzed in some detail with a focus on the question of exactly what role Grossmann played in it.
\end{abstract}

\section{Introduction}

The history of general relativity is a subject which has been written about extensively.\cite{RennJ2007Genesis} Nevertheless, most historical accounts of the emergence of the theory focus on Einstein's role in it, or at least they tell the story from a point of view that is largely defined by our view of Einstein's achievements. Indicative of this fact is the treatment of Marcel Grossmann's contribution to the emergence of general relativity in the secondary literature. Neither does a biographical account of his life and work exist nor has any attempt been made to analyze in some detail his particular contribution to the genesis of the theory, despite the fact that a big and successful international conference series is named after him. In this article, an attempt is made to fill these gaps to some extent. While a full-fledged biography is beyond the scope of the present article, an attempt is made to sketch at least Grossmann's biography and intellectual achievements by accounting for his published work. I also review the Einstein-Grossmann theory, which has been analyzed repeatedly in the literature and in great detail,\cite{PaisA1982Subtle,NortonJ1984Einstein,StachelJ1989Einstein,CPAE04,JanssenM1999Rotation,%
SauerT2005Paper,StraumannN2011Notebook,RaezT2013Applicability} but it will be done here specifically from Grossmann's perspective. The account draws mainly on published sources as well as on documents in the Albert Einstein Archives.

\section{Marcel Grossmann (1878--1936)}

Marcel Grossmann was born on 9 April 1878 in Budapest, Hungary.\cite{SaxerW1936Grossmann,KollrosL1937Grossmann,PaisA1982Subtle,CPAE01}
The son of a large machine-shop owner, he was the descendant of an old Swiss family, originating from H\"ongg, near Zurich. He spent his childhood and attended primary school in Budapest. At the age of 15 he returned
with his parents to Switzerland and attended secondary school in Basel. Apparently, he was a model student. A report card from April 1894 gave him the best grade (1) for effort in all subjects, and the best grade for achievement in almost all subjects, except for technical drawing (2), freehand drawing (2), and physical exercise (3).\footnote{See his \emph{Quartal-Zeugnis} of the \emph{Obere Realschule zu Basel}, extant in the Archives of the Library of the Swiss Federal Institute of Technology, Zurich, Switzerland, (ETH-Bibliothek, Hochschularchiv der ETH Z\"urich, in the following abbreviated as ETH Archives), call nr. Hs.421a:17). As explained on the card, ``1'' was the highest grade, ``5'' the lowest, concordant with most grading systems, until today, in use in Germany. Around that time, however, most Swiss schools and institutions changed to a grading system in which ``6'' is the highest grade, and ``1'' the lowest. The latter grading system was in use at the ETH already at the time of Grossmann's studies there.\label{note:reportcard}}

\begin{figure}[t]
\begin{center}
\includegraphics[width=8cm]{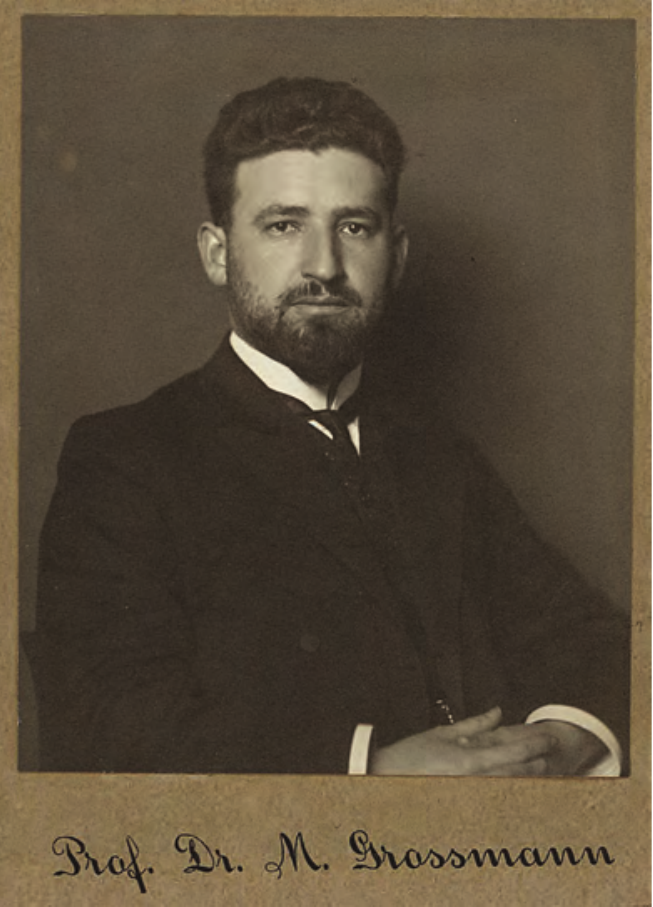}
\caption{Marcel Grossmann (1878--1936). \copyright ETH Bibliothek Z\"urich, Bild\-archiv.}
\label{fig:Grossmann1}
\end{center}
\end{figure}

After graduating from middle school, he entered the Swiss Polytechnic School in Zurich, now the Swiss Federal Institute of Technology (referred to as ETH in the following),
in its department VI, the School for
Mathematics and Science Teachers, in October 1896.\footnote{For the following, see Reference \citenumns[esp. pp.~43--44]{CPAE01}, as well as the \emph{Programme der eidgen\"ossischen polytechnischen Schule} for the years 1896--1900, and the \emph{Matrikel} kept for each ETH student at the ETH Archives.} The department was divided into two sections. Grossmann enrolled in section VI A, which was headed by Adolf Hurwitz (1859--1919) and specialized in mathematics, physics, and astronomy. The ETH counted a total of 841 students in 1896, but only 11 students enrolled in section VI A for the winter semester 1896/97. As is well-known, among Grossmann's peers of section VI A were Albert Einstein (1879--1955) and Mileva Mari\'c (1875--1948), the only woman in that class.
Two other students of Grossmann's entry class, Jakob Ehrat (1876--1960) and Louis Kollros (1878--1959), remained in Grossmann's cohort until the final examinations in 1900. Unlike in its five engineering school departments, in department VI there was no formal curriculum. Here the course of lectures to be taken by the students was determined on a more or less individual basis each year by the head of the department. Nevertheless, comparison of the students' transcripts shows that required classes were very much the same for all students in section VI A in the first two years. They included courses on calculus, analytic geometry, descriptive geometry, mechanics, projective geometry, and determinants in the first year; in the second year Grossmann and his fellow students took classes on differential equations, infinitesimal geometry, projective geometry, number theory, geometry of numbers, mechanics, physics, theory of scientific reasoning, Kant's critique of pure reason, geometric theory of invariants, complex analysis, potential theory, theory of definite integrals, and an introduction to celestial mechanics. In addition, the
students were free to take a choice of non-obligatory courses. The mandatory part of their schedule comprised some twenty hours a week. Not surprisingly, spending the better part of the week with a handful of peers in the same lecture room for two years, friendships were bound to build up. As is well-known, one such friendship soon developed between Grossmann and Einstein. They would hang out after school in Zurich's \emph{Caf\'e Metropol} and talk about ``everything that could be of interest to young folks with open eyes'' \citenumns[147]{EinsteinA1955Erinnerungen}.

\begin{figure}[t]
\begin{center}
\includegraphics[width=8cm]{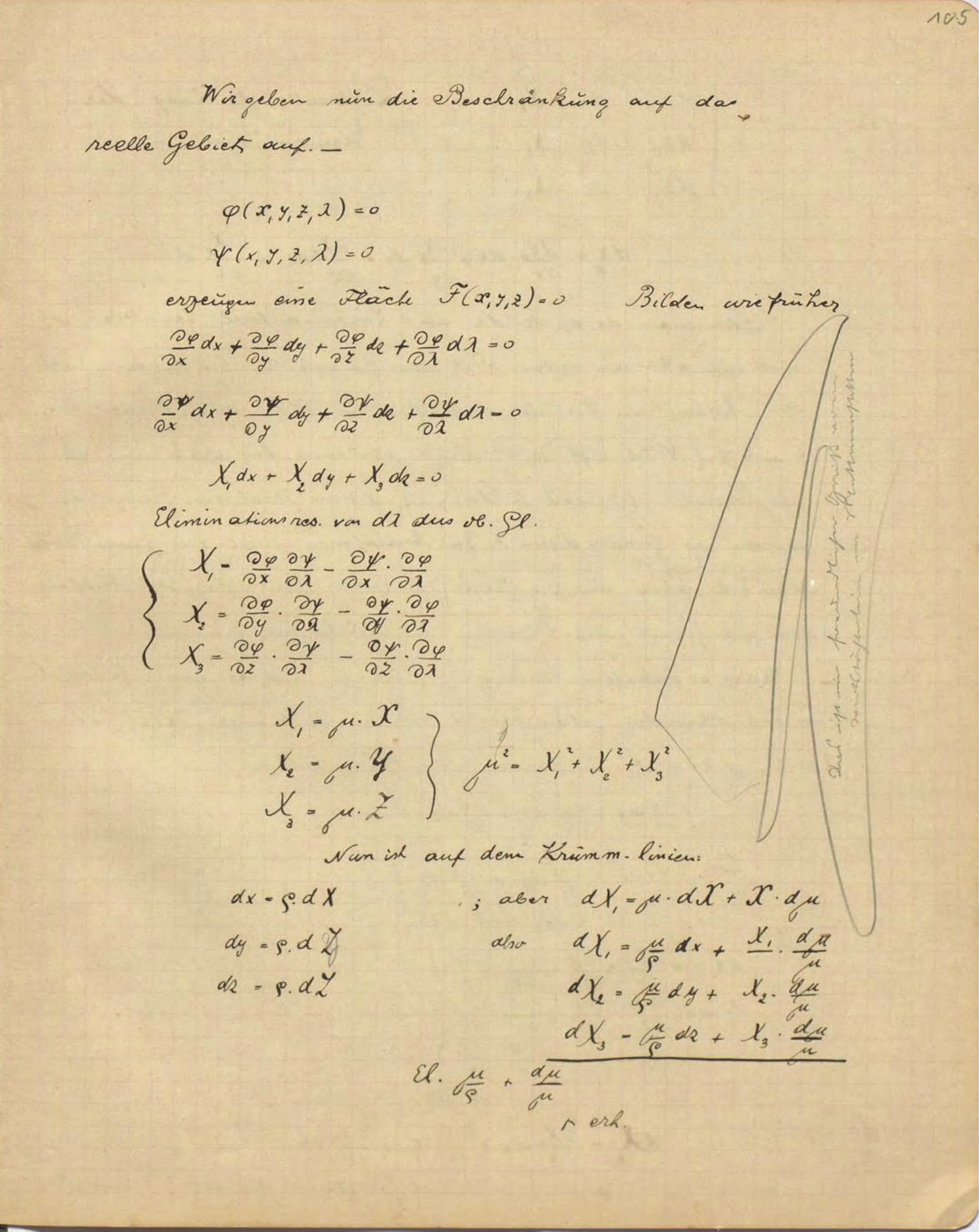}
\caption{A page from Grossmann's notes of Carl Friedrich Geiser's lectures on infinitesimal geometry, given in the summer term 1898 at the Swiss Polytechnic. This page shows a pencil comment probably made by Einstein. ETH archives, Hs 421:16, p.~105. Grossmann's lecture notes are available online at the ETH's e-manuscripta platform.}
\label{fig:GeiserNotes}
\end{center}
\end{figure}

In the third and fourth year, the curriculum of the students in section VI A diversified somewhat. Einstein and Mari\'c attended lectures in physics and spent more time in the laboratory, Grossmann, Ehrat, and Kollros continued their mathematical studies. Grossmann was a conscientious and industrious student. He took notes during the lectures and worked them out in bound booklets, carefully and
meticulously, in a clean and neat handwriting. Those notebooks, valuable witnesses of the mathematical knowledge and training of the day, were later given to the ETH archives and are accessible to the public.\footnote{ETH archives, call nr.\ Hs 421:10--33. Facsimiles of Grossmann's \emph{Ausarbeitungen} have recently been made accessible online by the ETH library on their e-manuscripta platform.} Einstein later recalled that Grossmann would lend him his \emph{Ausarbeitungen} so that he could use them to prepare for his examinations.\footnote{See Einstein to Grossmann, 15 March 1924 (AEA 11 105), \citenumns[16]{EinsteinA1949Autobiographisches}, \citenumns[147]{EinsteinA1955Erinnerungen}, see also Einstein to Mileva Mari\'c, 16 February 1898 \citenumns[Doc.~39]{CPAE01}. Indeed, in one of Grossmann's \emph{Ausarbeitungen} there is a marginal note that was made, in all probablity, by Einstein. On p.~105 of the third part of the \emph{Ausarbeitung} of a course by Carl Friedrich Geiser (1843--1934) on infinitesimal geometry (ETH archives, Hs 421:16, see Fig.~\ref{fig:GeiserNotes}), there is a couple of erratic pencil strokes across the page and next to it a note in neat German Gothic handwriting, saying: ``This is a friendly compliment by the little innkeeper's daughter in Mettmenstetten.'' (``Das ist ein freundlicher Gru{\ss} vom Wirtst\"ochterlein in Mettmenstetten.''). In August 1899, Einstein vacationed in Mettmenstetten, canton of Zurich \citenumns[374]{CPAE01}. \citenumns[212, note 5]{CPAE01} mentions that there are marginal notes in Grossmann's \emph{Ausarbeitungen} of \emph{Funktionentheorie} and \emph{Elliptische Funktionen} (ETH archives, Hs 421:19,20). Einstein's later claim that he skipped most of the
lectures at the ETH appears to be an urban legend, however. The rules of the ETH strictly required regular
attendance and several of Grossmann's and Einstein's classmates were reprimanded or even relegated for excessive absences.}

On 27 July 1900, final oral examinations took place for the section VI A class of 1896. The mathematicians Grossmann, Ehrat, and Kollros were examined in complex analysis, geometry, arithmetic and algebra, theoretical physics, and in astronomy. The physicists Einstein and Mari\'c were examined in theoretical physics, experimental physics, complex analysis, and in astronomy. For the mathematicians, grades in the
mathematical subjects were doubled. In addition, the grade for a diploma thesis was quadrupled for evaluation of the final grade. Grossmann, who had written a diploma thesis on non-Euclidean geometry with Wilhelm Fiedler, scored an average of 5.23, second in his class after Louis Kollros who scored 5.45. Ehrat scored an average of 5.14, Einstein 4.91, only Mari\'c failed the examination with a score of 4.0.

After obtaining his diploma, Grossmann obtained a position as \emph{Assistent} to Otto Wilhelm Fiedler (1832--1912), full professor for descriptive geometry and projective geometry at the ETH since 1867. This was a typical career step in an academic vita.\footnote{Of Grossmann's classmates, Ehrat also took an assistantship at the ETH after graduation, Einstein did not succeed in landing such a job despite multiple
applications. Kollros, who later became a professor at the ETH, first began work as an instructor at a Swiss gymnasium.} It allowed Grossmann to obtain his Ph.D., supervised by Fiedler, already in 1902 with a thesis ``On the Metric Properties of Collinear Structures''\cite{GrossmannM1902Eigenschaften}. The topic of the thesis was from the field of projective geometry, a field of expertise of Fiedler's, and the aim of the thesis was to give a detailed discussion of the focal and metric properties of collinear planes and bundles with a special emphasis on the aspect of projective duality. It also extended the concept of the characteristic of plane centric collineations to arbitrary collineations and, given two collinear spaces, to put this characteristic into relation to the coordinates of the collinear planes and bundles.\footnote{Einstein referred
to Grossmann's doctoral project in a letter to Mari\'c: ``Grossmann is getting his doctorate on a topic that is connected with Fiedlering [fiddling: non-translatable pun] and non-Euclidean geometry. I don't know exactly what it is.'' (``Grossmann doktoriert \"uber ein Thema, welches mit Fiedlerei und nichteuklidischer Geometrie zusammenh\"angt. Ich weiss nicht genau was es ist.'' \citenumns[Doc.~131]{CPAE01}) All of
Grossmann's classmates obtained a Ph.D.\ from the University of Zurich: Kollros and Einstein in 1905, Ehrat in 1906. The polytechnic school was only granted the privilege of awarding Ph.D.\ degrees in 1909.} Results from his thesis were published in a brief paper, which, however, appeared only in 1905.\cite{GrossmannM1905Eigenschaften}

\begin{figure}[t]
\begin{center}
\includegraphics[width=12cm]{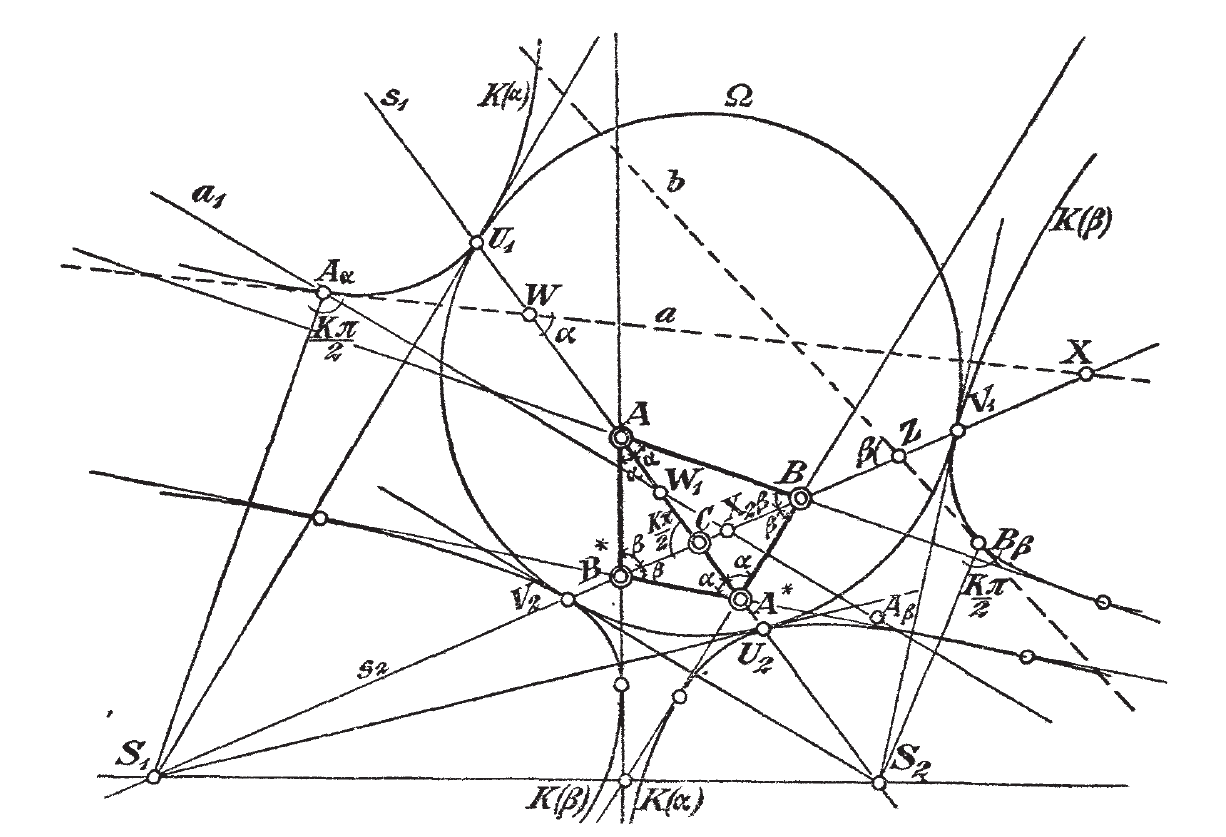}
\caption{Grossmann was an expert in synthetic constructions of non-Euclidean geometry in the framework of Cayley-Klein geometry. This figure (\citenumns[22]{GrossmannM1904Konstruktionen}, \citenumns[504]{GrossmannM1904Konstruktion}) illustrates the construction of a triangle with given right angle (at point $C$) and two arbitrary angles $\alpha$ and $\beta$ in a hyperbolic geometry represented by the conic section $\Omega$. The inside of the complex $\Omega$ represents the hyperbolic plane, its border the improper points at infinity. Straight lines which have a given angle $\alpha$ with $s_1$ or $\beta$ with $s_2$ are tangents to the conic sections $K(\alpha)$ or $K(\beta)$, and the task is to find a common tangent to both $K(\alpha)$ and $K(\beta)$, which is shown to be a problem only of second rather than of fourth degree.}
\label{fig:Konstruktion}
\end{center}
\end{figure}

On 31 August 1901, Grossmann was appointed to a position as an instructor at Thurgau Kantonsschule in Frauenfeld
\citenumns[316, note~2]{CPAE01}. During his tenure in Frauenfeld, he published a detailed account of what he called \emph{Fundamental Constructions of non-Euclidean Geometry}\cite{GrossmannM1904Konstruktionen}. An extract from that work appeared the same year also in \emph{Mathematische Annalen}.\cite{GrossmannM1904Konstruktion} With this work, which allegedly evoked praise by David Hilbert (1862--1943) \citenumns[323]{SaxerW1936Grossmann}, Grossmann established himself as an expert in non-Euclidean and projective geometry. The point of these investigations was to show that and how one can perform the elementary constructions of Euclidean geometry also for the case of non-Euclidean geometries, both hyperbolic and elliptic. Grossmann based his constructions on the concepts of Cayley-Klein geometry. In this framework,\cite{KowolG2009Geometrie} the set of improper, infinitely far away points of, say, two-dimensional hyperbolic geometry, is represented by a real, non-degenerate conic section $\Omega$ in the (Euclidean) plane, the so-called absolute conic section. Given such an absolute conic section, the hyperbolic plane is then formed by all points within $\Omega$. If $A$ and $B$ are two such points, and $U_1$ and $U_2$ the points of intersection between $\Omega$ and the straight line connecting $A$ and $B$, then the distance $r$ between $A$ and $B$ is given by the logarithm of the cross ratio
\begin{equation}
r = k \ln(U_1,U_2,A,B),
\end{equation}
where $k$ is a real constant. This Cayley-Klein metric now allows the construction of all elementary geometric objects by elementary geometric means, and Grossmann proceeds to show how the fundamental construction tasks of Euclidean geometry can now be transferred to the non-Euclidean case, see Fig.~\ref{fig:Konstruktion} for an illustration of his method.

In 1905, Grossmann moved to Basel to teach at his own former school, the \emph{Oberrealschule}. At that time, on the basis of his published research,
he also became \emph{Privatdozent} at the University of Basel, an unpaid academic title with the privilege (and obligation) to lecture at the university that was needed in order to be considered for a
professorship at a German language university. In Basel, Grossmann published two geometry textbooks, one on
analytic geometry\cite{GrossmannM1906GeometrieAna} and one on descriptive geometry\cite{GrossmannM1906GeometrieDar}. Generations of mathematics and engineering students, in Basel, at the ETH, and elsewhere, would learn geometry from these textbooks and its various later editions.\footnote{The textbook on analytic geometry saw a second
edition, revised by two colleagues of the Basel \emph{Oberrealschule} in 1914; the textbook on
descriptive geometry was reprinted in a second, revised edition in 1912, and in a third edition in 1917. An expanded two-volume version of the textbook on descriptive geometry was published in 1917, with a second edition in 1921/22 and a third edition in 1932, a textbook on descriptive geometry written especially for engineers appeared in 1927.}

When his academic teacher Fiedler asked for a leave due to ill health in 1906, Grossmann was asked to step in for him and to teach the course on descriptive geometry at the ETH in the winter semester of 1906/07.\footnote{ETH-Bibliothek, Archive, SR2: Pr\"asidialverf\"ugungen 1906, Pr\"asidialverf\"ugung Nr.~456 vom 15.10.1906 and Nr.~532 vom 23.11.1906.} Fiedler resigned for good
in June 1907 as of 1 October, his position was advertised, and Grossmann who was listed \emph{secundo loco} was appointed his successor on 22 July 1907, after Martin Disteli (1862--1923) in Dresden had rejected a call.\footnote{As his student and successor, Grossmann wrote an appreciative obituary after Fiedler's death on 19 November 1912.\cite{GrossmannM1913Fiedler}}  The initial appointment was for three years, but it was extended for another ten years in 1910, as was usual with such appointments.

If Grossmann's career from an ambitious student to a professor at the ETH was rather direct and without setbacks, his friend Einstein's career was less so. As is well-known,\cite{CPAE01} Einstein's attempts to obtain an assistantship anywhere failed flatly despite various letters of application. To make matters more
difficult, Einstein and his fellow student Mileva Mari\'c had fallen in love and, unfortunately, Mari\'c had become pregnant in the spring of 1901, an unplanned circumstance that surely contributed to her failing the final examinations again at her second attempt in summer 1901.

Einstein had also applied in July 1901 for the position at the Thurgau Kantons\-schule that Grossmann would obtain \citenumns[Doc.122]{CPAE01}. But Grossmann's father, Julius Grossmann (1843--1934), who was an old friend of Friedrich Haller (1844--1936), director of the Swiss Patent Office in Bern, recommended Einstein for a position at this office, and in June 1902 Einstein began to work there as a Technical Expert. Since Einstein now had a tenured job, he was able to marry Mileva Mari\'c in January 1903. Coincidentally, the same year, Grossmann got married to Anna Keller (1882--1967).\footnote{Einstein's and Grossmann's private lives continued to develop in parallel for some time. The following year, Grossmann's son Marcel Hans (1904--1986) was born 30 January, while Einstein's son Hans Albert (1904--1973) was born 14 May. In 1909, Marcel and Anna Grossmann had a daughter, Elsbeth Grossmann (1909--1986), while Einstein's second son Eduard (1910--1965) was born 28 July 1910.} When Einstein obtained his Ph.D. in 1905, he dedicated his doctoral thesis to his friend Grossmann.\cite{EinsteinA1905Bestimmung} A few years later, Einstein was considering his next career steps.
In 1908, he asked Grossmann for advice in applying for a teacher's position at a technical school in Winterthur \citenumns[Doc.~71]{CPAE05}. Nothing came of that, but shortly thereafter, he obtained his habilitation at the University of Bern, and, in 1909, he accepted an associate professorship at the University of Zurich. Although in close proximity again, little is known about Grossmann's and Einstein's interactions during the latter's tenure at Zurich University.

Grossmann, in the meantime, engaged himself in all aspects of mathematics: research, teaching, and disciplinary organization. In 1909, he showed how projective constructions of hyperbolic geometry can also be done using only a ruler, if a fixed auxiliary circle is given.\cite{GrossmannM1909Konstruktionen} He lectured about projective geometry and projective constructions to Swiss teachers.\cite{GrossmannM1909Aufbau} In 1910, he confirmed a conjecture in the field of photogrammetry formulated by Sebastian Finsterwalder (1862--1951) in 1897. Photogrammetry, in some sense the inverse of projective geometry, is concerned with the problem of constructing original three-dimensional objects, if only two-dimensional photographs of those objects are given. In a report on the state of the field presented to the \emph{Deutsche Mathematiker-Vereinigung}, Finsterwalder had claimed that given four photographs of an object, one can always construct the object uniquely up to a conformal scale factor but the explicit construction was practically unfeasible \citenumns[14--15]{FinsterwalderS1897Grundlagen}. In a lecture to the Zurich \emph{Naturforschende Gesellschaft}, Grossmann confirmed the conjecture about the practical impossibility of the actual construction by showing ``in a purely geometric way'' that the relevant surfaces would be given as a section between a developable surface of nineteenth degree with another surface of fifth degree and that after subtracting improper solutions, there exist 56 of those surfaces.\cite{GrossmannM1910Loesung}
In the same year, he co-founded the Swiss Mathematical Society, whose presidency he held in the years 1916 and 1917. An important early activity of the Society was the inauguration of a long-term editorial project publishing the writings of Leonhard Euler (1707--1783).\cite{NeuenschwanderE2010Jahre} In 1911, Grossmann published a
detailed fifty-page report on mathematics instruction at the ETH, as part of a broad report on mathematics
education in Switzerland, requested by the \emph{Commmission internationale de l'Enseignement math\'ematique}.\cite{GrossmannM1911Unterricht} Earlier, he had pushed for the establishment of holiday courses for mathematics teachers at Swiss gymnasia and middle schools.\cite{GrossmannM1909Referat} Similar courses had been established a few years earlier in Berlin and G\"ottingen and were intended to supplement
teachers' knowledge in the mathematical sciences \citenumns[146--147]{HallettMEtal2004Lectures}. When Einstein moved to Prague as a full professor in 1911, he was asked by Grossmann to lecture at one of those holiday courses, a request that Einstein felt unable to turn down \citenumns[Doc.~266]{CPAE05}. In a lecture to the Zurich \emph{Naturforschende Gesellschaft} on non-Euclidean geometry given of 29 January 1912, Grossmann also discussed the question of the geometric nature of real physical space.\cite{GrossmannM1912Geometrie}

With Grossmann now an important player in ETH's faculty, it is not surprising that he was also involved in bringing Einstein back to Zurich. In the winter term 1911/1912, Grossmann took over the chair of the department for mathematics and physics teachers, and one of the first things he must have done in this capacity is to write to Einstein informally to sound him out about his willingness to accept an offer to join the ETH faculty. Ever since Hermann Minkowski (1864--1909) had moved to G\"ottingen in 1902 the professorship for higher mathematics had been vacant, a situation that was lamented by Grossmann and his colleagues. When Einstein, in a letter of 18 November 1911 \citenumns[Doc.~307]{CPAE05}, indicated that he would consider coming back to his alma mater, the issue was discussed two weeks later at a meeting of the Swiss School Council and ETH's president Robert Gnehm (1852--1926) was asked to enter into formal negotiations with Einstein.\footnote{ETH archives, Schulratsprotokoll, meeting of 2 December 1911, nr.117.} Even after Gnehm had initiated his negotiations and when Einstein was receiving competing offers from the Netherlands, Grossmann continued to assist informally with advice as to how to best accelerate the process \citenumns[Docs.~319, 321]{CPAE05}. Grossmann must have been satisfied when, already on 30 January 1912, Einstein was appointed Professor of Theoretical Physics at the ETH effective 1 October 1912, despite some internal opposition from his colleagues \citenumns[Doc.~291]{CPAE05}. He had succeeded in securing a first-rate scholar for the science education of the ETH students and he could be looking forward to having his good friend around as a colleague again.

During Einstein's tenure at the ETH, he and Grossmann engaged in an intense and very successful collaboration, in which Einstein's physics training and Grossmann's mathematical background came together in a fruitful search for a relativistic theory of gravitation. The collaboration culminated in their joint publication of an ``Outline (German: \emph{Entwurf}) of a general theory of relativity and a theory of gravitation." This period and the collaboration will be discussed in more detail below. Here we will first continue to give an account of Grossmann's life and work.

Grossmann's intellectual biography after the \emph{Entwurf} episode continues to be dominated by his research, teaching, and administrative duties as professor of mathematics at the ETH. But during the war and for some time after the war years, Grossmann also engaged in patriotic activities. He wrote essays in the daily \emph{Neue Z\"urcher Zeitung}, published pamphlets and gave lectures, in which he emphasized the need to strengthen the national unity of the different parts of Switzerland.\cite{GrossmannM1915Erneuerung,GrossmannM1915Erneuerung2,GrossmannM1915Forderungen,%
GrossmannM1915Anregungen,GrossmannM1917Rolle}

Toward the end of the war and in the immediate postwar period, Grossmann engaged in the publication of a periodical. On 25 October 1918, an initiative committee issued an announcement, signed by Grossmann and two others, of the founding of a cooperative with the aim of publishing a new periodical called the \emph{Neue Schweizer Zeitung} and an invitation to possible subscribers and donors. Members of the cooperative had to be Swiss nationals. The first issue of the \emph{Neue Schweizer Zeitung} appeared on 20 December 1918, and it was then published twice weekly. The paper was intended to provide a forum for open debate of all issues concerning Swiss affairs, supporting the ideals of Swiss democracy, federalism, and national unity. As is clear from an invitation for subscribers, dated 27 December, Grossmann functioned as president of the board of management and as such was also a member of the editing committee. So was his brother Eugen Grossmann (1879--1963), a professor of economics at the University of Zurich. Both Marcel Grossmann and his brother regularly published contributions in the new paper. It continued to appear for three and a half years until 29 June 1922 when it stopped publication, and during this period Grossmann published some 40 contributions in the \emph{Neue Schweizer Zeitung}. A note, dated 20 June 1922, announcing the end of the paper due to economic problems was still signed by Grossmann as president of the board of management.

Another of Grossmann's activities in the immediate postwar years concerned an effort to reform the national Swiss regulations and requirements for obtaining a secondary-school diploma that would qualify for university studies, the so-called \emph{Maturit\"atszeugnis} or, short, \emph{Matura}.\cite{VonlanthenAEtal1978Maturitaet} The educational system in Switzerland has strictly been under authority of the cantons, the only influence that the federal government could take on the requirements of secondary school diplomata was through regulations for admission in the medical professions and through entrance requirements to the Swiss Federal Polytechnic (ETH) school in Zurich. Since a variety of different secondary schools, gymnasia, middle schools, business, professional and vocational schools existed in the various cantons with widely different curricula, standards, and examination rules, a committee was installed with the task of formulating rules for standards that would be acceptable nationwide. The debate and discussion process continued for several years and touched on several hotly disputed issues. One point of contention was the issue whether knowledge in the old languages Latin and Greek as it was taught in the traditional gymnasia should be required as it had been the case traditionally for students of the medical sciences. This issue collided with the wish that so-called \emph{Realgymnasien}, i.e., secondary schools which focussed on the sciences rather than on the languages, should be allowed to prepare for university studies on the basis of a science-centered curriculum. Also at stake were various different schooling traditions in the various cantons.

Mathematics was a core subject in all curricula and traditionally it was the ETH faculty themselves who decided on the admission of their students.%
\footnote{Remember that in 1895 Einstein had applied for admission at the ETH without any secondary-school leaving certificate and being two years under the regular admission age of eighteen. He did well in the science part of the exam, so that the ETH physics professor Heinrich Weber (1843--1912) gave him permission to attend his lectures, but otherwise Einstein failed the exam, and the ETH's director advised him to complete a regular curriculum at the Aargau cantonal school which would give him a qualifying \emph{Maturit\"atszeugnis}, see \citenumns[10--13]{CPAE01}.} \cite{GrossmannM1913Vorbildung}
As a professor of mathematics at the ETH, Grossmann was a member of the committee who were put in charge of formulating a proposal for reform, and he worked in this capacity quite seriously and passionately. In a number of public statements made at various occasions,\cite{GrossmannM1919Mittelschulreform,GrossmannM1921Maturitaetsreform,GrossmannM1922Maturitaetsreform1,
GrossmannM1922Maturitaetsreform2,GrossmannM1923Loesung,GrossmannM1923Reifeerklaerung} he argued for a reform that was guided by a spirit of both liberality and high scientific standard. He argued against overloading the required curricula by demanding too much knowledge of details in favor of furthering skills of independent judgment and study. He also advocated a proposal according to which the \emph{Realgymnasien} should be given the privilege of awarding a federally recognized \emph{Maturit\"atszeugnis} without requiring the knowledge of old languages. Quite in the spirit of his patriotic activities during the war, he also emphasized the need of educating students in the spirit of becoming independent, democratic, and patriotic citizens. The debate sometimes degenerated into veritable polemics during which Grossmann at one point accused his own former school, the \emph{Realgymnasium} in Basel, where he also had been on the faculty for two years, of an utter lack of scholarly standard.\cite{GrossmannM1923Reifeerklaerung} In a pointed polemic, Grossmann had said publicly that Basel had a very good Gymnasium but a very bad \emph{Realschule}. His point was that the \emph{Realschule} would not sufficiently weed out bad students, its gradings and examinations being too lax and friendly.The background for the polemic was that an alleged bad reputation of Basel's \emph{Realgymnasium} undermined Grossmann's argument for a science-based \emph{Maturit\"atszeugnis}.%
\footnote{In view of this polemics, it is an odd fact that only one particular report card for his studies of the year 1895 with its excellent grades is extant in the Grossmann files at the ETH archives, see note \ref{note:reportcard} on p.~\pageref{note:reportcard} above.} The final revision of the decree regulating a federal \emph{Maturit\"atspr\"ufung}, which was passed on 20 January 1925 proved to be a defeat in some points for
Grossmann's position. It introduced three types of \emph{Matura} based on old languages (A), new languages (B), and on sciences (C), but did not put the \emph{Matura} of type C on a par with the other two types,
because it required in addition extra Latin examinations for students who wished to enter medical school.

Grossmann's devotion as a teacher and pedagogue and his passion for educational policy is reflected in an essay \cite{GrossmannM1926Diagnosen} he wrote a year after that ``failed attempt'' at a reform of the Swiss educational system. In that essay, he reflected on his experiences as a teacher and examinator, having examined ``several thousands'' of his own students as well as having participated in ``several hundred'' examinations by others all over the country. In that essay, he again expressed his conviction that not only teaching in general but also
\begin{quote}
instruction in specific disciplines should [...] primarily develop competencies, should create a frame of mind that enables the young person at the end of his studies to swim about without a swimming coach and without swimming rings, even when the current would flow in unexpected directions.%
\footnote{\foreignlanguage{ngerman}{``Auch Fachunterricht sollte [...] in erster Linie \emph{F\"ahigkeiten} entwickeln, eine
Geistesverfassung schaffen, die den jungen Menschen am Ende seiner Studienzeit in Stand setzt, ohne Schwimmlehrer und ohne Schwimmg\"urtel zu schwimmen, auch wenn die Str\"omung nach ganz anderer Richtung gehen sollte, als der Student w\"ahnte.''} \citenumns[66]{GrossmannM1926Diagnosen}.}
\end{quote}
Quite similarly he expressed himself in a contribution to the \emph{Festschrift} for his colleague August Stodola (1859--1912),\cite{GrossmannM1929Fachbildung} which is another passionate plea for the enlightening role of science and technology in society and for the need of a sound education in these fields.

As far as his own pedagogical efforts are concerned---beyond his teaching at the ETH with various new editions of his textbook on descriptive geometry---he alerted teachers to the significance of projective geometry and its concept of improper elements at the level of secondary school education.\cite{GrossmannM1923Raumelemente} It should also be mentioned in this respect that he supervised four doctoral dissertations in the field of non-Euclidean geometry.\cite{KollrosL1937Grossmann}

In terms of his own mathematical research, Grossmann went back to his field of synthetic geometry.
In 1922, he presented a talk to the Swiss Mathematical Society on projective constructions of
elliptic geometries.\cite{GrossmannM1922Geometrie} In 1924, he discussed complete focal systems of plane algebraic curves, extending a definition of focal points for algebraic curves given first by Julius Pl\"ucker (1801--1868).\cite{GrossmannM1925Fokalsystem} A year later, he gave a detailed geometric discussion of the construction of the horopter, i.e.\ the set of points in space that, in geometrical vision, are imaged at corresponding points of the eye's retina, geometrically a third-order curve generated by two congruent line bundles corresponding to the fixation lines of the visual rays.\cite{GrossmannM1925Darstellung} In 1927,
Grossmann obtained a patent for ``improvements relating for the production of cams for looms.''%
 \footnote{Swiss Patent CH121538 (A), submitted on March 10, 1927, also GB286710 (A), with application date for the UK of February 3, 1928. Grossmann's affinity to and close ties with the engineering departments at the ETH is also obvious in his obituary of Rudolf Escher (1848--1921), who had been ETH's chair for mechanical technology for some four decades.\cite{GrossmannM1922Escher}} The point of the patent was to define principles for a machine that
 would allow a precise and accurate grinding of a specific part of mechanical looms. As he explained in a
 companion publication,\cite{GrossmannM1927Schlagexzenter} his invention arose from geometric insight.
He realized that the relative motion of certain cams for looms with rollers mounted on shafts adapted for angular movement would constitute an enveloping surface arising from a system of congruent surfaces, and pointed out that study of such enveloping surfaces has a long tradition in mathematics since Gaspard Monge (1746--1818).%
\footnote{After approval of the patent, Grossmann obtained a grant over 3'000 Francs from the Benno Rieter Fonds to explore a practical realization of his invention but in 1930 gave back 2'500 Francs because he could not pursue this research due to health problems. ETH-Bibliothek, Archive, SR2: Pr\"asidialverf\"ugungen, Pr\"asidialverf\"ugung Nr. 357 (1928) and Nr. 161 (1930).}
In 1930,
he published another short note on constructions of circles and conic sections in projective and non-Euclidean geometry.\cite{GrossmannM1930Darstellung} In his last scientific publication, to be discussed
below, Grossmann took issue with Einstein's teleparallel approach to a unified field theory of gravity and
electromagnetism.\cite{GrossmannM1931Fernparallelismus}

Grossmann's work in the twenties was severely hampered by symptoms of an uncurable illness. We know some details about Grossmann's condition from a letter that he wrote on 12 March 1927 to Einstein's friend Heinrich Zangger (1874--1954), the director of the institute for forensic medicine at the University of Zurich (AEA 40-059). In this letter, Grossmann reports that he had always been healthy until the year 1915, when he first had a fit of dizziness during a mountain hike and noted that he lost security of grip in his right hand during mountain climbing. In the summer of 1917, he began to drag his right leg and had disturbances of speech. Since that time he had seen countless doctors for symptoms of impediments of motion, mainly on his right side. In his letter, Grossmann asked Zangger for an opinion that would corroborate his own conjecture that those symptoms were caused by intoxication due to unhealthy conditions in the lecture halls during the war years and that therefore he would be eligible for professional disability compensation. Grossmann's condition, in any case, had gotten worse over the years. In the summer semester
1924, he had to take a leave from his teaching duties for health reasons.\footnote{ETH-Bibliothek, Archive, SR2: Pr\"asidialverf\"ugungen, Pr\"asidialverf\"ugung Nr. 143 vom 3. April 1924.} In the summer of 1925, Einstein visited Grossmann in Z\"urich and wrote to Zangger that he had a ``peculiar nervous ailment with palsy'' but found him better than he had expected.%
\footnote{\foreignlanguage{ngerman}{``Ich habe Grossmann besucht, der ein eigent\"umliches nerv\"oses Leiden mit L\"ahmungen hat, habe ihn aber besser gefunden als ich gef\"urchtet hatte.''} Einstein to Heinrich Zangger, 18 August 1925 \citenumns[Doc.~262]{SchulmannR2012Seelenverwandte}.} Another leave was granted for the winter term 1925/26. On 10 March 1926 he was granted a partial dispensation of his teaching duties for the summer term 1926 but a month later, he had to take a full leave for that semester, too; and he was also relieved from teaching duties in the winter term 1926/27. When he asked Zangger for his opinion, his retirement as a professor was imminent. Zangger wrote to Einstein that he did not believe in the causal nexus with the conditions during the war years but rather thought Grossmann was suffering from multiple sclerosis. Zangger's diagnosis appears to have been accurate and, in any case, Grossmann asked for an early retirement, which was granted to him on 28 May 1927, effective 1 October 1927.\footnote{ETH-Bibliothek, Archive, SR2: Pr\"asidialverf\"ugungen, Pr\"asidialverf\"ugung Nr. 272 vom 28. Mai 1927.}

Marcel Grossmann passed on 7 September 1936, at the age of 58.

\section{Grossmann's Collaboration with Albert Einstein}

Let us now go back to the time of the scientific collaboration between Grossmann and Einstein.
Einstein left Prague on 25 July 1912 and registered his change of residence to Zurich on 10 August. With a family of four, his sons being 8 and 2 years of age, it must have taken a few days to settle in. The Grossmann family may have been of help to the Einstein family but Marcel himself was busy preparing for a lecture at the 5th International Congress of Mathematicians which took place in Cambridge from 22--28 August 1912.\cite{GrossmannM1913Zentralprojektion} Just a few days later, from 8 to 11 September, the \emph{Schweizerische Naturforschende Gesellschaft} held their annual meeting in Altdorf, and Grossmann presented there a ``projective proof of Lobatchevsky's absolute parallel construction.''\cite{GrossmannM1912Beweis}

We know, however, that soon after arriving in Zurich, Einstein and Grossmann started a collaboration that would almost lead to the discovery of general relativity and that would, in any case, result in Grossmann's most well-known scholarly achievement. As Louis Kollros recalled in 1956, Einstein approached Grossmann for help, saying:
\begin{quote}
Grossmann, you have to help me, or else I'll go crazy!%
\footnote{\foreignlanguage{ngerman}{``Grossmann, Du mu{\ss}t mir helfen, sonst werd' ich verr\"uckt!''}\citenumns[278]{KollrosL1956Einstein}.}
\end{quote}
And, famously, on 29 October Einstein reported to Arnold Sommerfeld (1868--1951):
\begin{quote}
I am now working exclusively on the gravitation problem and believe that I can overcome all difficulties with the help of a mathematician friend of mine here. But one thing is certain: never before in my life have I troubled myself over anything so much, and I have gained enormous respect for mathematics, whose more subtle parts I considered until now, in my ignorance, as pure luxury!%
\footnote{\foreignlanguage{ngerman}{``Ich besch\"aftige mich jetzt ausschliesslich mit dem Gravitationsproblem und glaube nun mit Hilfe eines hiesigen befreundeten Mathematikers aller Schwierigkeiten Herr zu werden. Aber das eine ist sicher, dass ich mich im Leben noch nicht ann\"ahernd so geplagt habe, und dass ich grosse Hochachtung f\"ur die Mathematik eingefl\"osst bekommen habe, die ich bis jetzt in ihren subtileren Teilen in meiner
Einfalt f\"ur puren Luxus ansah!''} \citenumns[Doc.~421]{CPAE05}.}
\end{quote}

In order to properly assess Grossmann's contribution to the genesis of general relativity, it is
crucial to identify as concretely as possible the starting point for his collaboration with Einstein. Unfortunately, we can only speculate about how exactly their joint work took off. Therefore, it will be necessary to recapitulate briefly Einstein's efforts in generalizing special relativity up until his move to Zurich \citenumns[Vol.~1, 81--113]{RennJ2007Genesis}.

Einstein had made a first step of generalizing special relativity by formulating the equivalence hypothesis in 1907.\cite{EinsteinA1907Relativitaetsprinzip} In the following years, the problem had remained dormant with him until the summer of 1911 when he came back to the problem in Prague. At that time, he realized that one of the consequences of the heuristic assumption of a strict equivalence between constant linear acceleration and static homogeneous gravitation might actually be observable with the bending of star light grazing the limb of the sun during a solar eclipse.\cite{EinsteinA1911Einfluss} The crucial point was that the equivalence assumption implied that the velocity of light $c$ depend on the gravitational field, i.e.\ the constant $c$ became a spatially variable function $c=c(x)$. Specifically, Einstein deduced that the velocity of light would depend on the gravitational potential $\Phi(x)$ as
\begin{equation}
c = c_0\left(1+\frac{\Phi}{c^2}\right).
\end{equation}
In early 1912, Einstein was surprised by a paper by Max Abraham (1875--1922)\cite{AbrahamM1912Theorie} who claimed that this relation follows readily from postulating a generalization of Poisson's equation of the form
\begin{equation}
\frac{\partial^2\Phi}{\partial x^2} +
\frac{\partial^2\Phi}{\partial y^2} +
\frac{\partial^2\Phi}{\partial z^2} +
\frac{\partial^2\Phi}{\partial u^2} = 4\pi\gamma\nu
\end{equation}
with an imaginary time $u=ict$, a gravitational constant $\gamma$ and a mass density $\nu$, together with
equations of motion
\begin{equation}
\ddot{x} = -\frac{\partial\Phi}{\partial x}, 
\ddot{y} = -\frac{\partial\Phi}{\partial y},
\ddot{z} = -\frac{\partial\Phi}{\partial z},
\ddot{u} = -\frac{\partial\Phi}{\partial u},
\end{equation}
where the dots indicate differentiation of the coordinates of a material ``world point'' with respect to its proper time. Further reflection and correspondence, however, made it clear that Abraham's argument was not as straightforward as it may have seemed since his utilization of four-dimensional vector calculus became inconsistent with the assumption of a variable $c$. Abraham conceded to Einstein's criticism by restricting the light-cone relation to an infinitesimal line element\cite{AbrahamM1912Berichtigung}
\begin{equation}
ds^2 = dx^2 + dy^2 + dz^2 - c^2dt^2
\label{eq:staticmetric}
\end{equation}
with variable $c$.

Einstein, in any case, was pushed by this discussion to further consideration of the theory of static gravitation based on the equivalence hypothesis. In late February 1912, he published a paper\cite{EinsteinA1912Lichtgeschwindigkeit} on the topic, in which he also represented the propagation of light by means of an infinitesimal line element (\ref{eq:staticmetric}) and in which he suggested to generalize Poisson's equation
with a differential equation of the static gravitational field that he gave as
\begin{equation}
\Delta c = kc\rho
\label{eq:fieldeqstatic1}
\end{equation}
with gravitational constant $k$ and matter density $\rho$. Here $\Delta$ denotes the spatial, three-dimensional Laplace operator.

He soon found fault with his differential equation (\ref{eq:fieldeqstatic1}). Just a few weeks later, in late March, he submitted a second paper\cite{EinsteinA1912Theorie} with a modified differential equation, which he now gave as
\begin{equation}
\Delta c = k\biggl\{c\rho + \frac{1}{2k}\frac{\operatorname{grad}^2c}{c}\biggr\}.
\end{equation}
He interpreted the second term in brackets as an energy density of the gravitational field.

So far, Einstein had made use of the heuristics of the equivalence hypothesis only by considering constant linear acceleration. But already these considerations had shown him that he had to work with infinitesimal
line elements. He also had learned that the problem would probably involve non-linear differential equations, which were needed in order to properly take into account the energy density of the gravitational field itself. Nevertheless, the theory was still a scalar theory for a single function $c$ representing both the speed of light and the gravitational potential.

The next step was to look at stationary rotating coordinates. Although much less explicitly documented,
it is clear that transforming the line-element (\ref{eq:staticmetric}) to rotating Cartesian coordinates will produce mixed terms with a coefficient that involves the rotation frequency $\omega$. Interpreting the rotation field, represented by $\omega$, as a gravitational field may have induced Einstein, at some point, to consider a general
line element\footnote{Since the argument here is a historical one, I am keeping rather strictly to the original notation, in this case using subscript indices for coordinate differentials.}
\begin{equation}
ds^2 = \sum_{i,k=1}^4g_{ik}dx_idx_k
\label{eq:gen_lineelement}
\end{equation}
as the representation of a general gravitational field, i.e., one in which the coefficients $g_{ik}$ were not necessarily produced by a coordinate transformation away from the Minkowski line element but rather would represent a generic, independently given gravitational field. Einstein, in any case, at some point
saw the analogy between infinitesimal line elements occurring in his theory with the two-dimensional
line element of a curved surface in Gaussian surface theory, about which he had learned in Geiser's lectures at the ETH (see Fig.~\ref{fig:GeiserNotes}).

We don't know whether Einstein made the transition from a scalar theory to a gravitation theory based on the general line element (\ref{eq:gen_lineelement}) before he came to Zurich and before he began to talk with Grossmann about his problem. In later recollections, he reconstructed the beginning of their collaboration by posing to Grossmann a rather specific mathematical question. In 1955, he wrote:
\begin{quote}
The problem of gravitation was thus reduced to a purely mathematical one. Do differential
equations exist for the $g_{ik}$, which are invariant under non-linear coordinate
transformations? Differential equations of this kind and \emph{only} of this kind were
to be considered as field equations of the gravitational field. The law of motion of
material points was then given by the equation of the geodesic line.\\
With this problem in mind I visited my old friend Grossmann who in the meantime had become professor
of mathematics at the Swiss polytechnic. He at once caught fire, although as a mathematician he had a somewhat skeptical stance towards physics.\footnote{%
\foreignlanguage{ngerman}{``Das Problem der Gravitation war damit reduziert auf ein rein mathematisches.
Gibt es Differentialgleichungen f\"ur die $g_{ik}$, welche invariant sind
gegen\"uber nicht-linearen Koordinaten-Transformationen? Solche Differentialgleichungen
und \emph{nur} solche kamen als Feldgleichungen des Gravitationsfeldes in Betracht.
Das Bewegungsgesetz materieller Punkte war dann durch die Glei\-chung der geod\"atischen Linie
gegeben.\\
Mit dieser Aufgabe im Kopf suchte ich 1912 meinen alten Studienfreund Marcel Gro{\ss}mann
auf, der unterdessen Professor der Mathematik am eidgen\"ossischen Polytechnikum
geworden war. Er fing sofort Feuer, obwohl er der Physik gegen\"uber als echter
Mathematiker eine etwas skeptische Einstellung hatte.''}%
\citenumns[p.~151]{EinsteinA1955Erinnerungen}.}
\end{quote}
This recollection may not have been entirely accurate or, at least, it was probably too brief. It completely neglects another important step: the question of the proper representation of the gravitating mass-energy density. The transition from a scalar theory to a theory based on a multi-component object also
implied a transition from a scalar mass-energy density to a multi-component mathematical complex that involved momentum flow and stresses. We are purposely avoiding the modern term ``tensor'' here because, as we will see, the introduction of the tensor concept, as we are used to it now in the context of general relativity, may have been the first of Grossmann's contributions when he began his discussions  with Einstein.

\begin{figure}[t]
\begin{center}
\includegraphics[width=8cm]{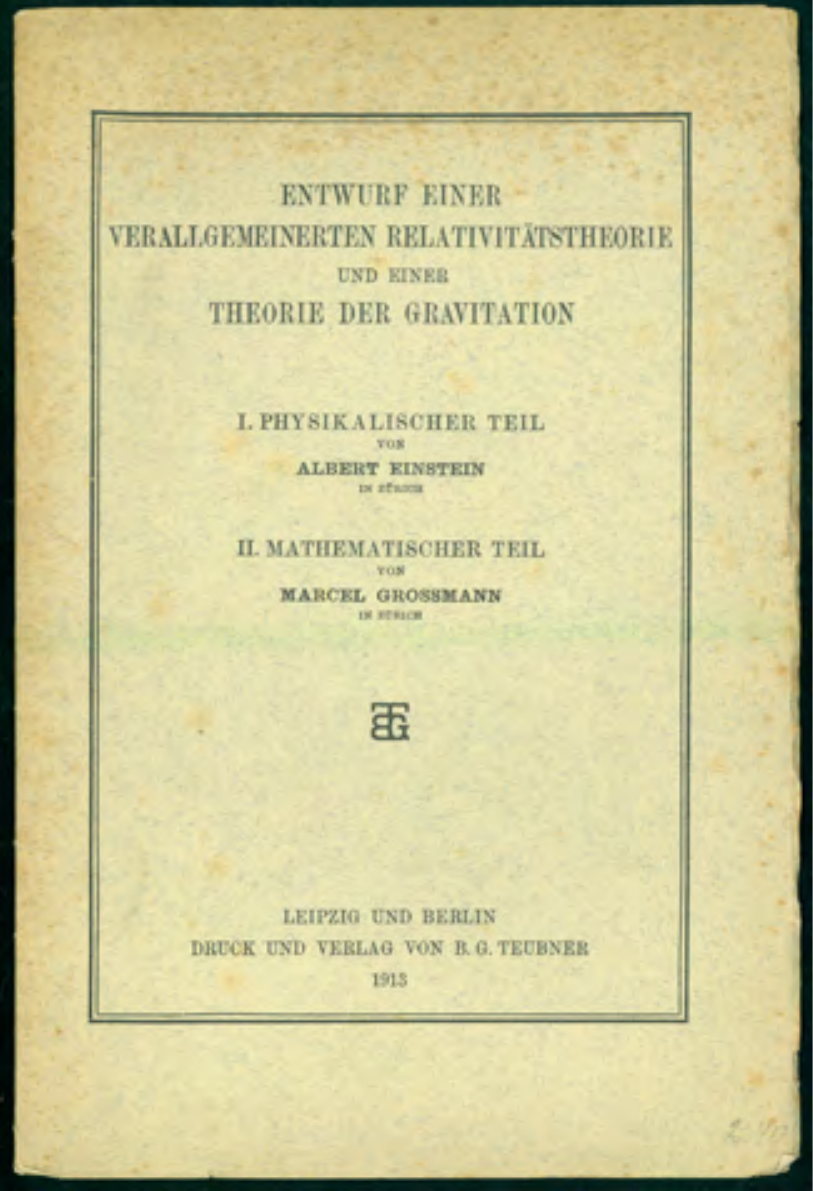}
\caption{Title page of the ``Outline of a Generalized Theory of Relativity and of a Theory of Gravitation'' published by Einstein and Grossmann in early summer 1913. This work, in which Einstein signed responsible for the `physical part' and Grossmann for the `mathematical part,' already contained all elements of the final theory of general relativity, except for the correct, generally covariant gravitational field equations. In particular, it contained an account of tensor calculus geared to the purposes of a relativistic theory of gravitation in Grossmann's mathematical part. The Einstein-Grossmann theory of this ``Outline'' was only given up by Einstein in the fall of 1915, when he succeeded in his final breakthrough to the general theory of relativity.}
\label{fig:Entwurf}
\end{center}
\end{figure}

There are two key documents, which give us insight into Grossmann's role in the collaboration with Einstein.\cite{NortonJ1984Einstein,RennJEtal1996Einstein,RennJ2007Genesis} The first document is Einstein's so-called Zurich Notebook \citenumns[Doc.~10]{CPAE04}, \citenumns[Vols.~1--2]{RennJ2007Genesis}. This is a bound notebook of some 85 written pages, 57 of which contain research notes, documenting the search for a relativistic theory of gravitation in the period between summer 1912 and spring 1913. All entries are in Einstein's hand but Grossmann's name appears twice in the notebook, at strategic places, as we will see below. The research documented in the Zurich Notebook leads directly up to the second document of relevance, a two-part paper published in the \emph{Zeitschrift f\"ur Mathematik und Physik} entitled ``Outline (\emph{Entwurf}) of a Generalized Theory of Relativity and of a Theory of Gravitation.''\cite{EinsteinAEtal1913EntwurfB} The work was
completed by mid-May 1913 and offprints (with independent pagination) were available before 25 June 1913.%
\footnote{See Einstein to V.~Vari\'cak, \citenumns[Doc.~Vol.~5,~439a]{CPAE10}. The presentation in \citenumns[Doc.~13]{CPAE04} refers to the offprint version as an independent publication that was later only ``reprinted'' in the journal. It was, however, a regular practice at the time that journal articles were made available as separately printed offprints (with independent pagination\citenumns[n.~74]{SauerT1999Relativity} and, possibly, extra title pages (see Fig.~\ref{fig:Entwurf}) before the relevant issue of the journal came out. The issue of the \emph{Entwurf} paper owned by the Swiss National Library (N11330/19 HelvRara) is of that kind, and it explicitly says on the back of the front page that it is a ``Sonderdruck'' from the journal.}
The paper was divided into two parts, a physical part, authored by Albert Einstein, and a
mathematical part, for which Grossmann signed responsible.

Taking clues from these two documents we can try to reconstruct Grossmann's contribution to the emergence of general relativity at this point in their collaboration. Large parts of the Zurich Notebook
may be reconstructed as the search for a gravitational field equation of the form \citenumns[p.~15]{EinsteinAEtal1913EntwurfB}, \citenumns[Vol.~1-2, 113--312, 489--714]{RennJ2007Genesis}
\begin{equation}
\Gamma_{\mu\nu} = \kappa \cdot \Theta_{\mu\nu}.
\label{eq:FieldEquationSchema}
\end{equation}
Here $\Gamma_{\mu\nu}$ stands for some operator acting on the metric coefficients $g_{\mu\nu}$ in a specific
way. Just what form $\Gamma_{\mu\nu}$ would have is the problem that Einstein and Grossmann were trying to solve. $\kappa$ denotes a constant that will be proportional to the gravitational constant, the
proportionality being determined on inspection of a limiting case in which the Poisson equation is being recovered. $\Theta_{\mu\nu}$ denotes the (contravariant) stress-energy-momentum tensor of matter and fields. Two comments are in order. First, as it stands, i.e., as long as $\Gamma_{\mu\nu}$ is not yet specified, Eq.~(\ref{eq:FieldEquationSchema}) is not a covariant equation, or rather, its covariance group is undetermined. Second,
although the equation was written with subscript indices, it was intended as a contravariant equation. The contravariant character of Eq.~(\ref{eq:FieldEquationSchema}) was expressed by the fact that Greek letters were used for the quantities $\Gamma$ and $\Theta$.

The notational peculiarities of the \emph{Entwurf} raise an important point regarding Grossmann's contribution to the emergence of the theory at this point. When Einstein approached Grossmann about the mathematics associated with the metric $g_{\mu\nu}$ it was not at all clear what status the object $g_{\mu\nu}$ actually had. In Gauss's surface theory, the three independent metric components $g_{11}$, $g_{12}$, $g_{22}$ of the two-dimensional line element were typically denoted by the letters $E$, $F$, $G$. The fact that the metric components are components of a \emph{tensor} is to be attributed to Grossmann. It was he who realized that a branch of mathematics had developed in which the Gaussian theory of surfaces featured only as a special example. Famously, Grossmann alerted Einstein to the existence of the so-called absolute differential calculus, which had been presented in a comprehensive joint paper by Gregorio Ricci-Curbastro (1853--1925) and Tullio Levi-Civita (1873--1941) in 1901.\cite{RicciGEtal1901Methodes,HermannR1975Paper,ReichK1994Entwicklung}

But Grossmann did more than simply find out about and exploit the absolute differential calculus for the purpose at hand. He realized very clearly that ``the vector analysis of Euclidean space in arbitrary curvilinear coordinates is formally identical to the vector analysis of an arbitrary manifold given by its line element.''\footnote{\foreignlanguage{ngerman}{``[...] die Vektoranalysis des auf beliebige krummlinige Koordinaten bezogenen euklidischen Raumes formal identisch ist mit der Vektoranalysis einer beliebigen, durch ihr Linienelement gegebenen Mannigfaltigkeit [...]''} \citenumns[p.23]{EinsteinAEtal1913EntwurfB}.}

Although in later recollections, Einstein credited Grossmann mainly for showing him the relevant literature, we must assume that Grossmann actually helped clarify the very mathematical status of the objects that
were entering the center stage of their theoretical efforts.

In any case, Grossmann gave a new and self-contained exposition of what he called ``general vector calculus'' (``allgemeine Vektoranalysis''). In doing so, Grossmann explicitly found it unnecessary to refer to any geometric concepts:
\begin{quote}
In doing this I deliberately did not draw on geometrical tools, as they contribute little to the illustration of the concepts of vector analysis.%
\footnote{\foreignlanguage{ngerman}{``Dabei habe ich mit Absicht geometrische Hilfsmittel beiseite gelassen, da sie meines Erachtens wenig zur Veranschaulichung der Begriffsbildungen der Vektoranalysis beitragen.''}\citenumns[p.~24]{EinsteinAEtal1913EntwurfB}.}
\end{quote}
In the \emph{Entwurf}, Grossmann proceeds to give an exposition of tensor calculus. He introduced covariant, contravariant, and mixed tensors for spaces of arbitrary dimensions and of any rank. The use of the word
``tensor'' in this context is a novelty.\footnote{As pointed out by John Stachel \citenumns[309--310]{NortonJ1992Content}, there is contemporary evidence that the usage of the term tensor for arbitrary $n$-dimensional co- and contravariant systems was first introduced in the \emph{Entwurf}. It is pointed out explicitly by Ernst Budde in his own 1914 exposition of three-dimensional calculus.\cite{BuddeE1914Tensoren} For a general historical account of the history of tensor calculus, see \citenumns{ReichK1994Entwicklung}.} Ricci and Levi-Civita had called these objects \emph{syst\`emes covariants ou contrevariants}, and they had never considered \emph{syst\`emes} of mixed transformation behavior, i.e.\ with a mix of covariant and contravariant indices. They had used superscripts and subscripts to indicate contravariant and covariant transformation behavior, except for coordinate differentials, which always carried subscript indices.\footnote{The inconsistency of writing coordinate differentials with subscript indices continued to be general practice for a few years after the establishment of final general relativity.} Grossmann introduced a notation where all indices were written as subscripts and the transformation character was indicated by writing the object itself with a Latin, Greek, or Gothic character for covariant, contravariant, or mixed tensors, respectively. In the \emph{Entwurf}, Grossmann defined tensor operations of a sum, external product, inner product (contraction), of changing covariant to contravariant objects by contraction with the fundamental tensor and vice versa (what we now call raising and lowering indices), and he introduced what we now call the trace of a (second-rank) tensor. With explicit reference to Elwin Bruno Christoffel, (1829--1900),\cite{ChristoffelEB1869Transformation}\footnote{With respect to the significance of Christoffel's work for Grossmann's elaboration of a tensor calculus for a relativistic theory of gravitation, it may be worth pointing out that from 1862--1869 Christoffel had been the founder and first director of the mathematical-physical department at the ETH and thus was Grossmann's predecessor in this function.} he also introduced covariant differentiation of a tensor, which he called ``expansion'' (``Erweiterung''). Next, he introduced a covariant concept of divergence by covariant differention of a tensor and contraction with the fundamental tensor, and he defined a generalized ``Laplacian operation'' as the combination of an expansion and a divergence.
He also paid some attention to the special case of antisymmetric tensors, a special case of which is the fully antisymmetric Levi-Civita tensor.

In the Zurich notebook, there is a page (05R) on which Einstein deduced an equation that we now recognize as the covariant divergence of the energy-momentum tensor.\footnote{Cf.~\citenumns[p.~383]{RennJ2007Genesis}. Note that the notation in the Zurich notebook at this point is still inconsistent with the notation introduced later in the \emph{Entwurf}. On the page at hand, the contravariant matter tensor is still expressed as $T_{ik}$, and the tensor density $\sqrt{-g}\,T_{ik}$ is denoted by $\Theta_{ik}$ instead.} Einstein there proceeds by looking a the Euler-Lagrange equations for a Hamiltonian $H=ds/dt$ and identifying the change of energy-momentum and the ponderomotive force density for a ``tensor of the motion of masses'' given as
\begin{equation}
\Theta_{ik} = \rho\frac{dx_i}{ds}\frac{dx_k}{ds},
\end{equation}
from which he obtained the energy-momentum balance equation
\begin{equation}
\sum_{\nu n}\frac{\partial}{\partial x_{n}}
\biggl(\sqrt{-g}\,g_{m\nu}\Theta_{\nu n}\biggr) -\frac{1}{2}\sum_{\mu\nu}\sqrt{-g}\,
\frac{\partial g_{\mu\nu}}{\partial x_{m}}\Theta_{\mu\nu} = 0,
\label{eq:energymomcons}
\end{equation}
a relation that we readily identify, in modern notation, as $(\sqrt{-g}\,{T^n}_m)_{;n}=0$. In the \emph{Entwurf}, Grossmann gives an explicit proof of the claim that the energy-momentum balance equation (\ref{eq:energymomcons}) is a generally covariant expression by showing that it is obtained as a covariant divergence of $\Theta_{\mu\nu}$.

Grossmann's insight that the energy-momentum balance equation is a perfectly valid, generally covariant relation should not be underestimated. It must have suggested that general covariance would be a viable goal if sophisticated concepts of advanced mathematics were made use of. It also showed that one half of the gravitation problem was already solved. Given a $g_{\mu\nu}$-field, the movement of matter is determined by a generally covariant equation of motion.

\begin{figure}[t]
\centering
\includegraphics[width=12cm]{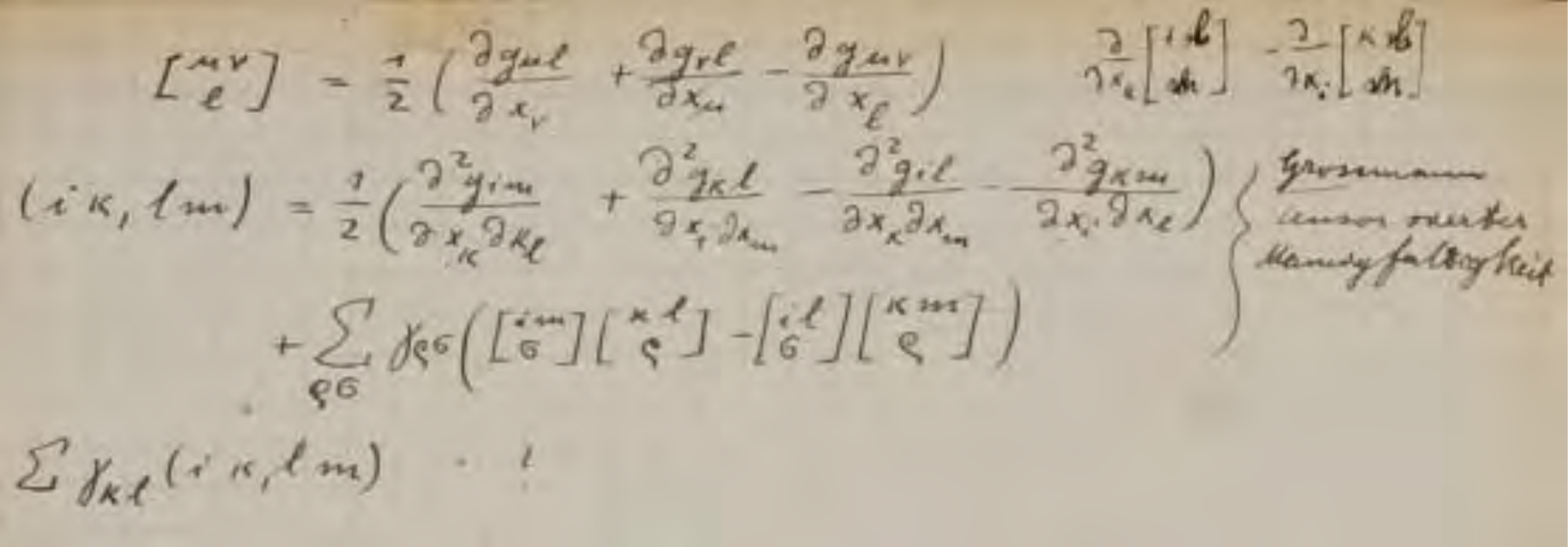}
\caption{Top portion of page 14L of Einstein's ``Zurich Notebook'' (AEA 3-006). This notebook documents the search for a generalized theory of relativity and a theory of gravitation during the period of collaboration between Grossmann and Einstein. This page shows that Grossmann introduced Einstein to the Riemann tensor as a mathematical resource for the general theory of relativity. \copyright The Hebrew University of Jerusalem, Albert Einstein Archives.}
\label{fig:ZN14L}
\end{figure}

Let us return now to the problem of finding a gravitational field equation. The schema of a field equation had to look like Eq.~(\ref{eq:FieldEquationSchema}) above. The problem was to find candidates for the gravitation tensor $\Gamma_{\mu\nu}$. There is a page in the Zurich Notebook where Grossmann's name appears right next to
the Riemann-Christoffel tensor, see Fig.~\ref{fig:ZN14L}. Apparently, Grossmann had shown Einstein the relevant object that would
open a path towards fully covariant gravitational field equations. Next to a definition of the
Christoffel symbols of the first kind
\begin{equation}
\begin{bmatrix} \mu\nu \\ l\end{bmatrix} =
\frac{1}{2} \biggl(
\frac{\partial g_{\mu l}}{\partial x_{\nu}} +
\frac{\partial g_{l\nu}}{\partial x_{\mu}} -
\frac{\partial g_{\mu\nu}}{\partial x_{l}}
\biggr)
\end{equation}
we find on that page the Riemann-Christoffel tensor in fully covariant form,
\begin{eqnarray}
R_{iklm} = (ik, lm) &= \displaystyle\biggl(
\frac{\partial^2g_{im}}{\partial x_k\partial x_l}
+ \frac{\partial^2g_{kl}}{\partial x_i\partial x_m}
- \frac{\partial^2g_{il}}{\partial x_k\partial x_m}
- \frac{\partial^2g_{km}}{\partial x_l\partial x_i} \biggr) \notag \\
&+ \displaystyle\sum_{\rho\sigma}\gamma_{\rho\sigma}\biggl(
\begin{bmatrix} im \\ \sigma\end{bmatrix}\begin{bmatrix} kl \\ \rho\end{bmatrix} -
\begin{bmatrix} il \\ \sigma\end{bmatrix}\begin{bmatrix} km \\ \rho\end{bmatrix}\biggr),
\label{eq:Riemanntensor}
\end{eqnarray}
next to the words: ``Grossmann tensor of fourth manifold.''
Since an object was needed that had only two free indices, Grossmann contracted the
Riemann tensor once. This amounted to the following operation:
\begin{equation}
G_{im} = \sum_{kl}\gamma_{kl} (ik,lm) = \sum_{k}(ik,km)
\label{eq:Riccitensor}
\end{equation}
and gave an expression that we now call the Ricci tensor.\footnote{Although, we do find the
Ricci tensor in a 1904 paper by Gregorio Ricci-Curbastro (1853--1925),\cite{RicciG1904Direzioni} it is not found in Ricci's and Levi-Civita's 1901
paper on the absolute differential calculus. It may therefore well be that Grossmann independently
invented the Ricci tensor at this point.}
The problem now was that if you form the Ricci tensor like this and you look at the
second derivative terms, you find the following terms
\begin{equation}
G_{im} = \sum_{k}\biggl(
\frac{\partial^2g_{im}}{\partial x_k^2}
+ \frac{\partial^2g_{kk}}{\partial x_i\partial x_m}
- \frac{\partial^2g_{ik}}{\partial x_k\partial x_m}
- \frac{\partial^2g_{km}}{\partial x_k\partial x_i} \biggr) + \dots
\label{eq:Gexpanded}
\end{equation}
In the limit that
\begin{equation}
g_{im} = \begin{pmatrix}1 & 0 & 0 & 0\\0 & 1 & 0 & 0 \\0 & 0 & 1 & 0\\0 & 0 & 0 & -c^2\end{pmatrix}
+h_{im} + \mathcal{O}(h_{ij}^2, (\partial h_{ij})^2)
\label{eq:weakfieldlimit}
\end{equation}
the first of the second derivative terms in (\ref{eq:Gexpanded}) reduces to the d'Alembertian ($x_4=ict$)
\begin{equation}
\Box \equiv \sum_{\mu=1}^4 \frac{\partial^2}{\partial x_{\mu}^2}
\end{equation}
but the other three second derivative terms do not vanish or take on simple forms which have
a reasonable physical interpretation. Einstein and Grossmann reasoned that these three terms
``should vanish''.

They also found a way to make these terms vanish. They imposed
a restrictive condition, let us call it the harmonic coordinate restriction, of the form
\begin{equation}
\sum_{kl} \gamma_{kl}\begin{bmatrix} kl \\ i\end{bmatrix} = \sum_{kl} \gamma_{kl}
\biggl(2\frac{\partial g_{il}}{\partial x_{k}} - \frac{\partial g_{kl}}{\partial x_{i}}\biggr) = 0.
\label{eq:harmonicrestriction}
\end{equation}
On p.~19L of the Zurich Notebook, it is shown that with this restriction, the Ricci tensor
reduces to
\begin{align}
2\tilde{G}_{im} = \sum_{kl}\biggl(\gamma_{k l}\frac{\partial^2g_{im}}{\partial x_{k}\partial x_l}
&- \frac{1}{2}\frac{\partial\gamma_{k l}}{\partial x_m}\frac{\partial g_{k l}}{\partial x_{i}}
+ \frac{\partial\gamma_{k l}}{\partial x_m}\frac{\partial g_{il}}{\partial x_{k}}
+ \frac{\partial\gamma_{k l}}{\partial x_i}\frac{\partial g_{mk}}{\partial x_{l}}\biggr)
\notag\\
&- \sum_{kl\rho\sigma}\biggl(\gamma_{\rho\sigma}\gamma_{k l}\frac{\partial g_{i\rho}}{\partial x_l}\frac{\partial g_{m\sigma}}{\partial x_{k}}
+ \gamma_{\rho\sigma}\gamma_{k l}\frac{\partial g_{il}}{\partial x_{\rho}}\frac{\partial g_{m\sigma}}{\partial x_{k}}\biggr).
\label{eq:Riccitensorred}
\end{align}
This expression is manifestly of the form that in the weak field limit (\ref{eq:weakfieldlimit})
it reduces to the d'Alembertian as
expected. In the Zurich Notebook, Einstein observed that this result was ``secure; valid for
coordinates that satisfy the Eq.~$\Delta\varphi = 0$.''

The trouble with this reasoning was that Einstein and Grossmann at this point were still
looking for a representation of $\Gamma_{\mu\nu}$. Since $G_{im}$ did not produce the correct limit
they were now considering $\tilde{G}_{im}$ as a candidate for $\Gamma_{\mu\nu}$. But the covariance group
of $\tilde{G}_{im}$ was restricted by the validity of the harmonic coordinate restriction.

Einstein therefore had to find a physics justification for the harmonic coordinate restriction. This proved
to be a fatal stumbling block for the expression $\tilde{G}_{im}$ as a candidate for $\Gamma_{\mu\nu}$. He briefly conjectured that the harmonic restriction (\ref{eq:harmonicrestriction}) would perhaps split
into the two conditions
\begin{equation}
\sum_{kl}\gamma_{kl}\frac{\partial g_{il}}{\partial x_k}
= -\sum_{kl} g_{il}\frac{\partial \gamma_{kl}}{\partial x_k} = 0
\label{eq:Hertzharmonic}
\end{equation}
and
\begin{equation}
\sum_{kl} \gamma_{kl}\frac{\partial g_{kl}}{\partial x_{i}} = 0.
\end{equation}
Looking at the weak field limit of these two conditions they found that the second condition amounted to the postulate that the trace of the weak field metric be constant, a condition clearly violated by the
static field metric (\ref{eq:staticmetric}). In an ad hoc move to remedy these difficulties, Einstein added a trace term to the weak field limit of the field equation, thus effectively writing down the weak field form of the final field equations of general relativity. But, alas, again Einstein found a problem in the interpretation of these equations when he confronted them with his further heuristic requirements.\cite{RennJEtal1999Heuristics,RennJ2007Genesis}

Yet, Grossmann showed Einstein a second way to get rid of the unwanted second derivative terms of $G_{im}$. %
\begin{figure}[t]
\centering
\includegraphics[width=12cm]{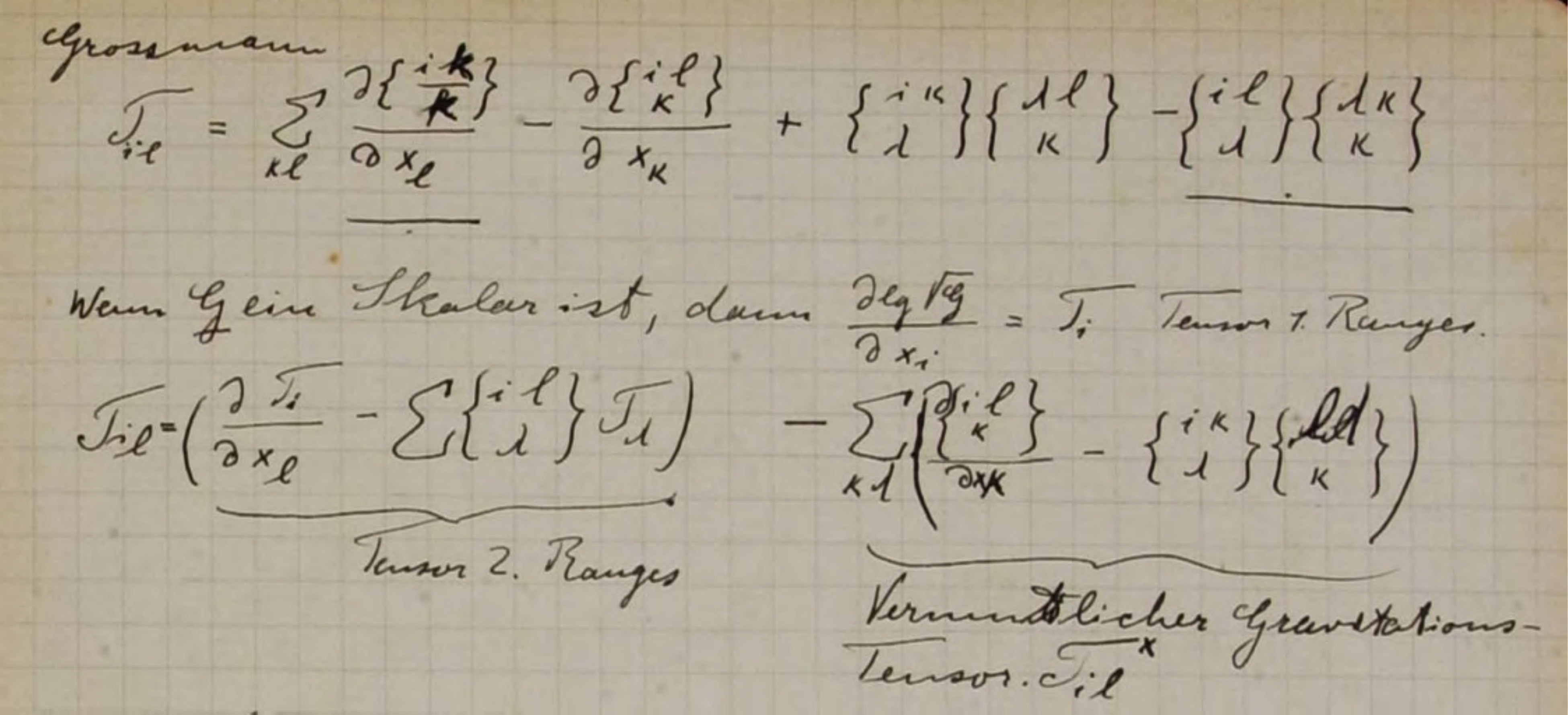}
\caption{Top portion of page 22R of the ``Zurich Notebook'' (AEA 3-006). Grossmann, at the time of their collaboration, also showed Einstein a way to extract a reduced quantity from the Riemann tensor that transforms as a tensor under unimodular coordinate transformations. Although given up at the time, this particular tensor reappeared three years later as a candidate gravitation tensor in the first of Einstein's famous four memoirs of November 1915 which mark the breakthrough to the final theory of general relativity. \copyright The Hebrew University of Jerusalem, Albert Einstein Archives.}
\label{fig:ZN22R}
\end{figure}
On p.~22R of the Zurich Notebook, we find Grossmann's name again next to a candidate gravitation tensor \citenumns[p.~451]{RennJ2007Genesis}, see Fig.~\ref{fig:ZN22R}.
The strategy was the same. This time the restrictive condition to be imposed on the Ricci tensor was that the determinant of the metric
transforms as a scalar. In addition, a condition similar to (\ref{eq:Hertzharmonic}) was assumed to hold.
Grossmann told Einstein to write the covariant Ricci tensor in the following form (see Fig.~\ref{fig:ZN22R})
\begin{equation}
G_{il} = \sum_{k}\underline{\frac{\partial\left\{\begin{matrix}ik\\k\end{matrix}\right\}}{\partial x_l}}
- \frac{\partial\left\{\begin{matrix}il\\k\end{matrix}\right\}}{\partial x_k}
+ \sum_{k\lambda}\left\{\begin{matrix}ik\\\lambda\end{matrix}\right\}\left\{\begin{matrix}\lambda l\\
k\end{matrix}\right\}
- \underline{\left\{\begin{matrix}il\\\lambda\end{matrix}\right\}\left\{\begin{matrix}\lambda k\\k\end{matrix}\right\}},
\label{eq:Gunderlined}
\end{equation}
where
\begin{equation}
\left\{\begin{matrix}ik\\\lambda\end{matrix}\right\}
= \sum_{\mu}\gamma_{\lambda\mu}\begin{bmatrix} ik \\ \mu\end{bmatrix}
\end{equation}
are the Christoffel symbols of the second kind.
Since
\begin{equation}
T_i\equiv\sum_{k}\left\{\begin{matrix}ik\\k\end{matrix}\right\} = \frac{1}{\sqrt{-g}}\frac{\partial\ln\sqrt{-g}}{\partial x_i}
\end{equation}
Grossmann argued that if $g$ transforms as a scalar, then $T_i$ transforms as a vector (a ``tensor of
1$^{\rm st}$rank''), and hence the underlined terms in (\ref{eq:Gunderlined}) represent the covariant derivative of a
vector and therefore transform as a second-rank tensor under unimodular transformations. This means
that the remaining two terms in (\ref{eq:Gunderlined})
\begin{equation}
G_{il}^{\ast} \equiv -\sum_{k}
\frac{\partial}{\partial x_k}\left\{\begin{matrix}il\\k\end{matrix}\right\}
+ \sum_{k\lambda}
\left\{\begin{matrix}ik\\\lambda\end{matrix}\right\}\left\{\begin{matrix}\lambda l\\k\end{matrix}\right\}
\label{eq:Novembertensor}
\end{equation}
also transform as a second-rank tensor under this restricted group of coordinate transformations.
Expanding the Christoffel symbols, we find that
\begin{equation}
G^{\ast}_{il} = \sum_{\kappa \alpha}\frac{1}{2}\frac{\partial}{\partial x_{\kappa}}
\left(\gamma_{\kappa\alpha}\left(
\frac{\partial g_{i\alpha}}{\partial x_l}
+ \frac{\partial g_{l\alpha}}{\partial x_i}
- \frac{\partial g_{il}}{\partial x_{\alpha}}
\right)\right)
+ \sum_{\kappa \lambda} \left\{\begin{matrix}i\kappa\\\lambda\end{matrix}\right\}\left\{\begin{matrix}\lambda l\\\kappa\end{matrix}\right\}.
\end{equation}
Assuming further that the condition
\begin{equation}
\sum_{\kappa}\frac{\partial \gamma_{\kappa\alpha}}{\partial x_{\kappa}} \equiv 0
\label{eq:Hertz}
\end{equation}
holds, and using
\begin{equation}
\sum_{\alpha}\gamma_{\kappa\alpha}\frac{\partial g_{i\alpha}}{\partial x_l} =
- \sum_{\alpha} g_{i\alpha}\frac{\partial\gamma_{\kappa\alpha}}{\partial x_l}
\end{equation}
the contravariant metric can be pulled outside the derivative and $G^{\ast}_{il}$ turns
into
\begin{equation}
\tilde{G}^{\ast}_{il} = \frac{1}{2}\sum_{\kappa}\gamma_{\kappa\alpha}\frac{\partial^2g_{il}}{\partial x_{\kappa}\partial x_{\alpha}}
+ \sum_{\kappa\lambda}\left\{\begin{matrix}i\kappa\\\lambda\end{matrix}\right\}\left\{\begin{matrix}\lambda l\\\kappa\end{matrix}\right\}.
\label{eq:Novembertensorred}
\end{equation}
This again was of the desired form of a single second derivative term which reduces to the
d'Alembertian for weak fields and to the Laplacian for weak static fields plus terms quadratic
in the derivatives of the metric which vanish in the weak field limit. But, again, the
derivation of this reduced ``gravitation tensor'' came at the cost of stipulating two additional
restrictive conditions, the unimodularity condition and condition (\ref{eq:Hertz}). But, again,
the physical interpretation of these two restrictions proved impossible for Einstein and
Grossmann, and so they discarded this approach as well.

In the Zurich Notebook, the same strategy of deriving a gravitation tensor of the form
\begin{equation}
\Gamma_{il} = \sum_{\alpha\beta}\gamma_{\kappa\alpha}\frac{\partial^2g_{il}}{\partial x_{\alpha}\partial x_{\beta}}
+ \mathcal{O}(\partial g)^2,
\label{eq:expectedform}
\end{equation}
was explored some more with yet different restrictive conditions but none of those attempts
proved feasible, and it is unclear whether Grossmann had his hand in any of the more outlandish
attempts along these lines that are recorded in the Zurich Notebook.

When the ``mathematical'' strategy of exploring the Riemann tensor as a resource for a derivation
of a suitable gravitation tensor $\Gamma_{\mu\nu}$ was exhausted, Einstein and Grossmann,
in a move of reflection, altered their strategy. All along, their first criterion in evaluating
the feasibility of candidate gravitation tensors was whether it was of the
form (\ref{eq:expectedform}), where the
terms quadratic in the first derivatives were to be determined by suitably restricting the
Riemann tensor for their needs.

Another heuristic requirement for the desired field equations arose from the covariant divergence
equation for the stress-energy tensor (\ref{eq:energymomcons}). This relation implied a requirement on the field equations because together with a field equation of the form (\ref{eq:FieldEquationSchema}) it implied
\begin{equation}
\left(\sqrt{-g}\,\Gamma^{\mu\nu}\right)_{;\nu} = 0,
\label{eq:Bianchicontracted}
\end{equation}
as indeed it does in the final theory where $\Gamma_{\mu\nu}$ is instantiated by the Einstein tensor.
Today the contracted Bianchi identity (\ref{eq:Bianchicontracted}) gives
a direct hint as to what the gravitation tensor should be. But remember, the Ricci tensor $G_{il}$ and its various reductions $\tilde{G}_{il}$, $G^{\ast}_{il}$, $\tilde{G}^{\ast}_{il}$ (cp.~Eqs.~(\ref{eq:Riccitensor}), (\ref{eq:Riccitensorred}), (\ref{eq:Novembertensor}), (\ref{eq:Novembertensorred})), or
the Einstein tensor, for that matter,\footnote{To be sure, the Einstein tensor had never been written down explicitly. But implicitly a trace term had been added to the reduced Ricci tensor $\tilde{G}_{il}$, which had then been considered as a candidate gravitation tensor, meaning that, effectively, the linearized Einstein equations had been considered---and dismissed \citenumns[vol.~1, 216--217; vol.~2, 632--637]{RennJ2007Genesis}.} had already been excluded for reasons of violating the expected
behavior in the weak static limit. The energy-momentum conservation (\ref{eq:energymomcons})
implied a heuristic requirement on the gravitational field equations because its individual terms were
interpreted realistically. Einstein argued that in the absence of any gravitational field, the
relation reduces to the divergence relation
\begin{equation}
{T^{\mu\nu}}_{;\nu} = 0.
\end{equation}
Consequently, the term
\begin{equation}
\frac{1}{2}\sum\sqrt{-g}\,\frac{\partial g_{\mu\nu}}{\partial x_{\sigma}}\,\Theta_{\mu\nu}
\end{equation}
was interpreted as the gravitational force density exerted on the ponderable matter $\Theta_{\mu\nu}$ by the gravitational field 
${\partial g_{\mu\nu}}/{\partial x_{\sigma}}$ .
This implied that on using the field equation (\ref{eq:FieldEquationSchema}) the term
\begin{equation}
\frac{1}{2}\sum\sqrt{-g}\,\frac{\partial g_{\mu\nu}}{\partial x_{\sigma}}\Gamma_{\mu\nu}
\end{equation}
had to be expressed as a coordinate divergence in order to guarantee that energy and momentum
conservation not be violated. The reasoning here was to some extent by analogy.

In electrostatics, Einstein argued, the momentum transferred onto ponderable matter of charge density $\rho$ by an electric field asscoiated with an electrostatic potential $\varphi$ is expressed by
$-{\partial\varphi}/{\partial x_{\nu}}$. The Poisson equation,
\begin{equation}
\Delta\varphi = \rho,
\end{equation}
guarantees conservation of momentum by virtue of the identity
\begin{equation}
\frac{\partial\varphi}{\partial x_{\nu}}\sum_{\mu}\frac{\partial^2\varphi}{\partial x_{\mu}^2}
= \sum_{\mu}\frac{\partial}{\partial x_{\mu}}
\left(
\frac{\partial\varphi}{\partial x_{\nu}}\frac{\partial\varphi}{\partial x_{\mu}}
\right)
-\frac{\partial}{\partial x_{\nu}}
\left(
\frac{1}{2}\sum_{\mu}
\left(\frac{\partial\varphi}{\partial x_{\mu}}\right)^2
\right).
\end{equation}
By analogy, Einstein and Grossmann now tried to construct the first derivative terms in such a way
that a relation of the form
\begin{align}
\text{Sum of }&\text{differential quotients} \notag \\
= &\frac{1}{2}\sum_{\mu\nu}\sqrt{-g}\,
\frac{\partial g_{\mu\nu}}{\partial x_{\sigma}}
\biggl\{\sum_{\alpha\beta}\frac{\partial}{\partial x_{\alpha}}
\left(
\gamma_{\alpha\beta}\frac{\partial\gamma_{\mu\nu}}{\partial x_{\beta}}
\right) \notag\\
&+ \text{further terms which vanish with the first approximation}\biggr\}.
\label{eq:heuristicschema}
\end{align}
While the heuristics of this reasoning was based on analogy, Grossmann provided Einstein with
a rigorous identity. He took the expression
\begin{equation}
U = \sum_{\alpha\beta\mu\nu}\frac{\partial g_{\mu\nu}}{\partial x_{\sigma}}
\frac{\partial}{\partial x_{\alpha}}
\left(
\sqrt{-g}\,\gamma_{\alpha\beta}\frac{\partial\gamma_{\mu\nu}}{\partial x_{\beta}}
\right)
\end{equation}
and transformed it, using partial integration and the relations
$\sqrt{-g}_{,\sigma}=(1/2)\sqrt{-g}\,g^{ik}g_{ik,\sigma}$
and
${g^{rs}}_{,l} = -g^{r\rho}g^{s\sigma}g_{\rho\sigma,l}$,
to show that the following identity holds
\begin{align}
\sum_{\alpha\beta\tau\rho}&\frac{\partial}{\partial x_{\alpha}}
\biggl(\sqrt{-g}\,\gamma_{\alpha\beta}
\frac{\partial\gamma_{\tau\rho}}{\partial x_{\beta}}\frac{\partial g_{\tau\rho}}{\partial x_{\sigma}}
\biggr)
-\frac{1}{2}\sum_{\alpha\beta\tau\rho}\frac{\partial}{\partial x_{\sigma}}
\biggl(\sqrt{-g}\,\gamma_{\alpha\beta}
\frac{\partial\gamma_{\tau\rho}}{\partial x_{\alpha}}\frac{\partial g_{\tau\rho}}{\partial x_{\beta}}
\biggr) \notag \\
=&\sum_{\mu\nu}\sqrt{-g}\,\frac{\partial g_{\mu\nu}}{\partial x_{\sigma}}
\biggl\{\sum_{\alpha\beta}\frac{1}{\sqrt{-g}}\frac{\partial}{\partial x_{\alpha}}
\biggl(\gamma_{\alpha\beta}\sqrt{-g}\,\frac{\partial\gamma_{\mu\nu}}{\partial x_{\beta}}\biggr)
-\sum_{\alpha\beta\tau\rho}\gamma_{\alpha\beta}g_{\tau\rho}
\frac{\partial\gamma_{\mu\tau}}{\partial x_{\alpha}}\frac{\partial\gamma_{\tau\rho}}{\partial x_{\beta}}
\notag\\
& + \frac{1}{2}\sum_{\alpha\beta\tau\rho}\gamma_{\alpha\mu}\gamma_{\beta\nu}
\frac{\partial g_{\tau\rho}}{\partial x_{\alpha}}\frac{\partial\gamma_{\tau\rho}}{\partial x_{\beta}}
- \frac{1}{4}\sum_{\alpha\beta\tau\rho}\gamma_{\mu\nu}\gamma_{\alpha\beta}
\frac{\partial g_{\tau\rho}}{\partial x_{\alpha}}\frac{\partial\gamma_{\tau\rho}}{\partial x_{\beta}}
\biggr\}.
\label{eq:Grossmannsidentity}
\end{align}
In order to interpret this identity, Einstein and Grossmann introduced a differential operator
\begin{equation}
\Delta_{\mu\nu}(\gamma) = \sum_{\alpha\beta}\frac{1}{\sqrt{-g}}\cdot
\frac{\partial}{\partial x_{\alpha}}\biggl(\gamma_{\alpha\beta}\sqrt{-g}\cdot
\frac{\partial \gamma_{\mu\nu}}{\partial x_{\beta}}\biggr)
-\sum_{\alpha\beta\tau\rho}\gamma_{\alpha\beta}g_{\tau\rho}
\frac{\partial\gamma_{\mu\tau}}{\partial x_{\alpha}}\frac{\partial\gamma_{\nu\rho}}{\partial x_{\beta}}
\label{eq:defDelta}
\end{equation}
and a gravitational stress-energy tensor
\begin{equation}
-2\kappa\cdot \vartheta_{\mu\nu} = \sum_{\alpha\beta\tau\rho}
\biggl(
\gamma_{\alpha\mu}\gamma_{\beta\nu}
\frac{\partial g_{\tau\rho}}{\partial x_{\alpha}}\frac{\partial\gamma_{\tau\rho}}{\partial x_{\beta}}
-\frac{1}{2}\gamma_{\mu\nu}\gamma_{\alpha\beta}
\frac{\partial g_{\tau\rho}}{\partial x_{\alpha}}\frac{\partial\gamma_{\tau\rho}}{\partial x_{\beta}}
\biggr)
\label{eq:defvartheta}
\end{equation}
and rewrote Grossmann's identity (\ref{eq:Grossmannsidentity}) in the form
\begin{equation}
\sum_{\mu\nu}\frac{\partial}{\partial x_{\nu}}
\left(
\sqrt{-g}\,g_{\sigma\mu}\vartheta_{\mu\nu}
\right)
-\frac{1}{2}\sum_{\mu\nu}\sqrt{-g}\,\frac{\partial g_{\mu\nu}}{\partial x_{\sigma}}
\vartheta_{\mu\nu} 
= -\frac{1}{2\kappa}\sum_{\mu\nu}\sqrt{-g}\,\frac{\partial g_{\mu\nu}}{\partial x_{\sigma}}
\Delta_{\mu\nu}(\gamma).
\label{eq:Grossmannsidentitycompact}
\end{equation}
By comparison with the conservation equation for matter (\ref{eq:energymomcons}),
they concluded that the quantity
$\vartheta_{\mu\nu}$ played the role of gravitational stress-energy, and by comparison of (\ref{eq:Grossmannsidentitycompact})
with their heuristic equation (\ref{eq:heuristicschema}),
they concluded that the gravitation tensor $\Gamma_{\mu\nu}$
entering the field equations (\ref{eq:FieldEquationSchema})
reads
\begin{equation}
\Gamma_{\mu\nu} = \Delta_{\mu\nu}(\gamma) - \kappa \cdot \vartheta_{\mu\nu},
\end{equation}
which renders the gravitational field equations in the form
\begin{equation}
\Delta_{\gamma\mu}(\gamma) = \kappa \left(\Theta_{\mu\nu} + \vartheta_{\mu\nu}\right).
\label{eq:Entwurfeq}
\end{equation}
Eqs.~(\ref{eq:defDelta}), (\ref{eq:defvartheta}), and (\ref{eq:Entwurfeq}) are the gravitational field equations of the Einstein-Grossmann theory in their contravariant form.

From (\ref{eq:energymomcons}) and (\ref{eq:Grossmannsidentitycompact}) it also follows that
\begin{equation}
\sum_{\mu\nu}\frac{\partial}{\partial x_{\nu}}
\left\{
\sqrt{-g}\,g_{\sigma\mu}
\left(
\Theta_{\mu\nu} + \vartheta_{\mu\nu}
\right)
\right\}
=0,
\label{eq:energycons}
\end{equation}
a relation expressing, according to Einstein, the validity of conservation laws for the union of
matter and gravitational field.

In a few months of their collaboration, Einstein and Grossmann had succeeded in formulating a relativistic theory of gravitation, which employed an adapted version of Ricci's and Levi-Civita's absolute differential calculus and which was generally covariant in all its parts, except
for the gravitational field equations. In the course of their joint work, they had been taking into consideration as candidate gravitation tensors the right mathematical objects and had even considered a linearized version of the final field equations of gravitation.

After their joint work of the \emph{Entwurf}\cite{EinsteinAEtal1913EntwurfB} was available in offprint form (see Fig.~\ref{fig:Entwurf}), Einstein and Grossmann engaged in various
activities to advertise and promote their results.
\clearpage

On 9 September 1913, they presented their theory to the 96th annual meeting
of the \emph{Schweizerische Naturforschende Gesellschaft}, which took place that year in Frauenfeld.
Again, they presented the work
with the same division of labor and responsibility. Abstracts of their presentations were
published in the society's \emph{Verhandlungen},\cite{EinsteinA1913Gravitationstheorie,GrossmannM1913Begriffsbildungen}
and longer versions were published in the \emph{Vierteljahrsschrift} of the \emph{Naturforschende
Gesellschaft in Z\"urich} \cite{EinsteinA1913Grundlagen,GrossmannM1913Begriffsbildung}.
French translations of the latter pieces appeared in
\emph{Archives des sciences physiques et naturelles}.\cite{EinsteinA1914Bases,GrossmannM1914Definitions}
Einstein alone presented the work two weeks later, on 23 September 1913,
at the 85th meeting of the \emph{Gesellschaft Deutscher Naturforscher und \"Arzte} in Vienna.
An account of his report was published under his name in the \emph{Physikalische Zeitschrift}.\cite{EinsteinA1913Stande}
Discussion remarks following his presentation were also published, but a few
weeks later, Einstein addressed again a comment by Hans Rei{\ss}ner (1874--1967), which he felt he had not satisfactorily answered.\cite{EinsteinA1914Antwort} He also defended the Einstein-Grossmann theory against criticism by Gustav Mie (1868--1957).\cite{EinsteinA1914Prinzipielles}

In its issue of 30 January 1914, the \emph{Entwurf} was eventually printed as a regular article in the
\emph{Zeitschrift f\"ur Mathematik und Physik},\cite{EinsteinAEtal1913EntwurfZ} a journal edited by the
applied mathematicians Carl Runge (1856--1927) and Rudolf Mehmke (1857--1944). The journal print is important for its addendum, signed only by Einstein, which pointed out
two recent advances in the understanding of the original \emph{Entwurf} theory \citenumns[Doc.~26]{CPAE04}. For one, Einstein here formulated
the infamous ``hole argument'' (``Lochbetrachtung'') by means of which he had convinced himself
that generally covariant gravitational field equations were not compatible with basic assumptions
on causality and the postulate that the metric components are determined uniquely by the matter tensor
$\Theta_{\mu\nu}$.%
\footnote{The hole argument played a significant role in recent debates in the philosophy of space and
time, see \citenumns{NortonJD2011Hole} and references therein.}
He also gave a version of the basic \emph{Entwurf} equations in terms of mixed tensor densities. Introducing
the quantities
\begin{align}
\mathfrak{T}_{\sigma\nu} &= \sum_{\mu}\sqrt{-g}\,g_{\sigma\mu}\Theta_{\mu\nu}, \\
\mathfrak{t}_{\sigma\nu} &= \sum_{\mu}\sqrt{-g}\,g_{\sigma\mu}\vartheta_{\mu\nu},
\end{align}
the conservation laws (\ref{eq:energymomcons}) and (\ref{eq:energycons}) can be written in the (deceptively) simple form as
\begin{equation}
\sum_{\nu}\frac{\partial\mathfrak{T}_{\sigma\nu}}{\partial x_{\nu}}
=\frac{1}{2}\sum_{\mu\nu\tau}\frac{\partial g_{\mu\nu}}{\partial x_{\sigma}}
\gamma_{\mu\tau}\mathfrak{T}_{\tau\nu}
\label{eq:addendumI}
\end{equation}
and
\begin{equation}
\sum_{\nu}\frac{\partial}{\partial x_{\nu}}
\left(
\mathfrak{T}_{\sigma\nu} + \mathfrak{t}_{\sigma\nu}
\right) = 0,
\label{eq:addendumII}
\end{equation}
respectively, and the \emph{Entwurf} field equations turn into
\begin{equation}
\sum_{\alpha\beta\mu}\frac{\partial}{\partial x_{\alpha}}
\left(
\sqrt{-g}\, \gamma_{\alpha\beta}g_{\sigma\mu}\frac{\partial\gamma_{\mu\nu}}{\partial x_{\beta}}
\right)
=\kappa \left( \mathfrak{T}_{\sigma\nu} + \mathfrak{t}_{\sigma\nu} \right).
\label{eq:addendumIII}
\end{equation}

Einstein also continued work on the gravitation problem in a collaboration with the Dutch physicist
Adriaan Fokker (1887--1972).
In a joint paper, received by \emph{Annalen der Physik} on 19 February 1914,
they reinterpreted a scalar gravitation
theory by the Finnish physicist Gunnar Nordstr\"om (1881--1923) using the tools of the absolute differential calculus and compared the Nordstr\"om theory
to the Einstein-Grossmann theory.\cite{EinsteinAEtal1914Gravitationstheorie}
That comparison was then also subject in a presentation that
Einstein gave on 9 February 1914 to the \emph{Naturforschende Gesellschaft} in Zurich.
Grossmann was present during Einstein's presentation but was active only as a participant in the discussion.\cite{EinsteinA1914Theorie}

Einstein left Zurich on 21 March 1914 \citenumns[p.~636]{CPAE05} and took up his new position as member
of the Prussian Academy in early April, after a short visit with Paul Ehrenfest (1880--1943) in Leyden. Einstein's move
to Berlin put an end to his collaboration with Grossmann. But presumably in the final weeks
before leaving Zurich, he and Grossmann penned another joint publication,%
\footnote{Strictly speaking, this is the \emph{only} joint paper, considering the fact
that the original \emph{Entwurf}\cite{EinsteinAEtal1913EntwurfB}
as well as the direct derivatives of it\cite{EinsteinA1913Gravitationstheorie,GrossmannM1913Begriffsbildungen,%
EinsteinA1913Grundlagen,GrossmannM1913Begriffsbildung,EinsteinAEtal1913EntwurfZ,%
EinsteinA1914Bases,GrossmannM1914Definitions} were published with clearly delineated responsibilities by Einstein and Grossmann for their ``physical'' and ``mathematical'' parts, respectively.}
which was published on 29 May 1914 also in \emph{Zeitschrift f\"ur Mathematik und Physik}.\cite{EinsteinAEtal1914Kovarianzeigenschaften}

The starting point of their second joint paper was the insight that Einstein had formulated in the
addendum to the journal print of the \emph{Entwurf}. By way of introduction, they recapitulated the
achievement of the earlier \emph{Entwurf}. That theory, they pointed out, contained two kinds of equations.
The first kind were generalizations of equations in the special theory of relativity, which govern the behavior of matter or material processes in general, for a given
gravitational field. These equations had been shown to be generally covariant. They had also established
a set of equations that determined the gravitational field if the quantities that determine the material processes are considered as given. This equation was a generalization of Poisson's equation and there
was no special relativistic analog to it. They had not been able to determine the covariance group
of those equations. It was only known that they would be covariant under linear transformations but it had
remained unclear which further coordinate transformations would maintain the form of those field
equations. The purpose of their second note then was to determine the maximum covariance group for the
\emph{Entwurf} field equations.

The first paragraph gave the basic equations of the theory in mixed tensor density form, just as
Einstein had done in the addendum, see Eqs. (\ref{eq:addendumI}), (\ref{eq:addendumII}), and (\ref{eq:addendumIII}) above. The second paragraph reiterated the hole argument. Although first formulated by Einstein, it was here presented in their joint paper as a ``proof that if a solution for the $\gamma_{\mu\nu}$ for given $\Theta_{\mu\nu}$ is already known, then the general covariance of the equations allows for the existence of further solutions'' \citenumns[p.~218]{EinsteinAEtal1914Kovarianzeigenschaften}. This mathematical fact would imply
that ``a complete determination of the fundamental tensor $\gamma_{\mu\nu}$ of a gravitational field with given $\Theta_{\mu\nu}$
by a generally-covariant system of equations is impossible'' \citenumns[p.~217]{EinsteinAEtal1914Kovarianzeigenschaften}. The hole argument, just as earlier Grossmann's
identity (\ref{eq:Grossmannsidentity}), is a central tenet of the Einstein-Grossmann theory, and it seems that Grossmann was concerned
about the invariant-theoretic consequences that it implied. The proof proceeds like this.
Consider a region (the ``hole'') in four-dimensional space-time in which there are no material processes, i.e., $\Theta_{\mu\nu}=0$. Assume that the metric field $\gamma_{\mu\nu}(x)$ is uniquely determined by the given field of $\Theta_{\mu\nu}(x)$, also in the matter-free region. Now introduce new coordinates $x'$ such that the new coordinates agree with the old ones outside the matter free region and only differ inside it.
Such a coordinate transformation would produce a transformed metric field $\gamma_{\mu\nu}'(x')$ inside
the matter-free region, but leave $\Theta_{\mu\nu}$ invariant across the entire space-time since
outside the hole we have $x' = x$ and inside we have $\Theta_{\mu\nu}' = 0 = \Theta_{\mu\nu}$.
Generally covariant field equations then would allow to have $\gamma_{\mu\nu}'(x)$ also as a solution and
hence one obtains two different metric fields $\gamma_{\mu\nu}'$, $\gamma_{\mu\nu}$ compatible with one and the same matter field $\Theta_{\mu\nu}$.

The version of the hole argument presented in their joint paper essentially reiterated the version that
Einstein had given earlier in the addendum. But the presentation in their joint paper immediately
proceeds to correct an erroneous argument of the addendum. Einstein had earlier believed that
Eq.~(\ref{eq:addendumII}) is necessarily only covariant under linear coordinate transformations. Einstein and Grossmann
now pointed out that this conclusion only holds ``if one assigns tensorial character to the quantities $\mathfrak{t}_{\mu\nu}/\sqrt{-g}$ which, it turned out, is not justified'' \citenumns[p.~218]{EinsteinAEtal1914Kovarianzeigenschaften}. Plugging Eq.~(\ref{eq:addendumII}) into (\ref{eq:addendumIII}), they instead now argued
that
\begin{equation}
B_{\sigma} \equiv \sum_{\alpha\beta\mu\nu}
\frac{\partial^2}{\partial x_{\nu}\partial x_{\alpha}}
\biggl(
\sqrt{-g}\,\gamma_{\alpha\beta}g_{\sigma\mu}\frac{\partial\gamma_{\mu\nu}}{\partial x_{\beta}}
\biggr) = 0
\label{eq:defB}
\end{equation}
represents a ``real restriction on the choice of coordinate systems'' \citenumns[p.~219]{EinsteinAEtal1914Kovarianzeigenschaften}.

The bulk of the paper was devoted to giving a variational derivation of the \emph{Entwurf} field equations,
which would allow them to make some inferences about the invariant-theoretic properties
of their theory. They showed that the field equation could be represented by the variational
principle
\begin{equation}
\int\biggl(\delta H-2\kappa\sum_{\mu\nu}\sqrt{-g}\,T_{\mu\nu}\delta\gamma_{\mu\nu}\biggr)d\tau = 0
\end{equation}
with a gravitational Lagrangian%
\footnote{In the terminology of the day, the principle was called ''Hamilton's principle'' and the Lagrangian was called ``Hamilton's function'', which is why the function was denoted by ``H''.}
\begin{equation}
H = \frac{1}{2}\sqrt{-g}\,\sum_{\alpha\beta\tau\rho}\gamma_{\alpha\beta}
\frac{\partial g_{\tau\rho}}{\partial x_{\alpha}}\frac{\partial \gamma_{\tau\rho}}{\partial x_{\beta}}.
\end{equation}
The core argument of their second paper consists of a proof of the invariance of the variational integral
$\int Hd\tau$ under general coordinate transformations that only respect the restrictive condition (\ref{eq:defB}). They interpreted their result to the effect that the gravitational field equations
possess the maximal covariance group that is compatible with the hole argument.

Their second joint paper only appeared when Einstein had already moved from Zurich to Berlin to take up
his position as member of the Prussian Academy.%
\footnote{No separately printed offprints of this article appear to be extant.}
With Einstein's absence from Zurich the
collaboration between the two friends effectively came to an end. There is only one more thing that
Grossmann contributed to the \emph{Entwurf} theory. Already from Berlin, but before 10 April 1914,
Einstein wrote to Paul Ehrenfest:
\begin{quote}
Grossmann wrote me that now he also is succeeding in deriving the gravitation equations from
the general theory of covariants. This would be a nice addition to our examination.%
\footnote{\foreignlanguage{ngerman}{``Grossmann schrieb mir, dass es ihm nun auch gelinge, die Gravitationsgleichungen aus der allgemeinen Kovariantentheorie abzuleiten. Es w\"are dies eine h\"ubsche Erg\"anzung zu unserer
Untersuchung.''}
\citenumns[Doc.~2]{CPAE08}.}
\end{quote}
Grossmann's letter, unfortunately, appears to have been lost, and we do not have any other evidence
of Grossmann's result. Apparently, he had found a way to recover the \emph{Entwurf} field equations (\ref{eq:Entwurfeq}) or (\ref{eq:addendumIII}) from the Riemann-Christoffel tensor (\ref{eq:Riemanntensor})
using the restrictive condition (\ref{eq:defB}).

To recapitulate: Grossmann's contribution to the \emph{Entwurf} theory consisted in the following.
\begin{itemize}
\item He clarified the mathematical foundation of the theory based on a general line element (\ref{eq:gen_lineelement}) and generalized the concept of a tensor
to a structure of $n-$th rank in $m$-dimensional space.
\item He identified the absolute differential calculus by Ricci and Levi-Civita as the relevant
mathematical toolbox for the problem of a relativistic theory of gravitation and transformed it
into a tensor calculus both with respect to notation and by generalizing it to mixed tensor densities.
\item He proved that the conservation law for matter (\ref{eq:energymomcons}) was a generally
covariant equation
by showing that it expresses the covariant divergence of $\Theta_{\mu\nu}$.
\item He identified the Riemann tensor as a relevant and rich resource for the problem of
constructing generally covariant gravitational field equations, and he showed Einstein several
ways of extracting a second rank object from the Riemann tensor that would have the desired
limiting form in the case of weak static fields.
\item After the failure of the mathematical strategy of constructing a field equation from the
Riemann tensor, he proved the central identity (\ref{eq:Grossmannsidentity}) from which the gravitational field equations
of the \emph{Entwurf} theory were read off.
\item In joint work with Einstein, he showed how the Einstein-Grossmann theory can be formulated in terms
of a variational principle and clarified its transformational properties in light of the hole argument.
\end{itemize}

A few months after Einstein had moved to Berlin the war broke out, a political course of events
that contributed to putting an end to the active collaboration between Einstein and Grossmann, even
if their friendship was not affected by the political turmoil. In any case, it was Einstein who
continued to work on the gravitation problem. In the fall of 1914, he wrote a first comprehensive
review paper of the Einstein-Grossmann theory, in which he also gave a new exposition of the relevant
mathematics.\cite{EinsteinA1914Grundlage} The review begins with giving credit to Grossmann's contribution. Einstein wrote:
\begin{quote}
In recent years, I have worked, in part together with my friend Grossmann, on a generalization of the
theory of relativity.%
\footnote{\foreignlanguage{ngerman}{``In den letzten Jahren habe ich, zum Teil zusammen mit meinem Freunde Grossmann, eine Verallgemeinerung
der Relativit\"atstheorie ausgearbeitet.''}%
\citenumns[Doc.~9, p.~73]{CPAE06}.}
\end{quote}
In the review Einstein referred to the Einstein-Grossmann theory as a ``general theory of relativity'' for the first time in the title, rather than calling it a ``generalized theory,'' as it appeared in the title of the \emph{Entwurf}. It was the Einstein-Grossmann theory as presented in this review that Einstein defended against criticism by the
mathematician Tullio Levi-Civita himself, and which he also defended in a course of lectures held in
the summer of 1915 to the mathematicians and physicists in G\"ottingen.

In the summer of 1915, plans were also under way to prepare a new edition of the collection of papers on the ``relativity principle'' first edited in 1913 by Otto Blumenthal (1876--1944).\cite{EinsteinAEtal1913Relativitaetsprinzip} Apparently, Arnold Sommerfeld who had initiated the first edition of the anthology and who also had contributed some annotation to Minkowski's paper in it, had asked Einstein which of his later works on relativity should be included in an augmented second edition. In his reply, Einstein mentioned the 1914 review paper but also suggested that he preferred to have none of the recent papers included since none of the expositions of the ``general theory of relativity'' would be complete and he intended to write a new self-contained presentation anyway. In that context, Einstein also commented on his collaboration with this friend:
\begin{quote}
Grossmann will never lay claim to being co-discoverer. He only helped in guiding me through the mathematical literature but contributed nothing of substance to the results.%
\footnote{\foreignlanguage{ngerman}{``Grossmann wird niemals darauf Anspruch machen, als Mitentdecker zu gelten. Er half mir nur
bei der Orientierung \"uber die mathematische Literatur, trug aber materiell nichts zu den
Ergebnissen [be]i.''}%
\citenumns[Doc.~96]{CPAE08} The second edition appeared in 1915 without any changes.}
\end{quote}
As is well-known, the breakthrough to general covariance occurred only a few months after Einstein's visit to G\"ottingen, and was documented in a series of four memoirs\cite{EinsteinA1915Relativitaetstheorie,EinsteinA1915Nachtrag,EinsteinA1915Erklaerung,EinsteinA1915Feldgleichungen} presented to the Prussian Academy, in which
Einstein regained general covariance,\cite{EinsteinA1915Relativitaetstheorie,EinsteinA1915Nachtrag} succeeded in the computation of Mercury's anomalous perihelion advance,\cite{EinsteinA1915Erklaerung} and finally completed his general theory of relativity by publication of the Einstein equations.\cite{EinsteinA1915Feldgleichungen} In the introductory paragraph of the first of those November papers, in which he proposed a theory of gravitation based on what we have called the
``November tensor''\citenumns[Vol. 1, p.192]{RennJ2007Genesis}, i.e., Eq.~(\ref{eq:Novembertensor}) above, covariant under general unimodular transformations, Einstein again mentioned his collaboration with Grossmann. He wrote:
\begin{quote}
Thus I came back to the postulate of a more general covariance of the field equations, which I had given up three years ago only with a heavy heart, when I worked together with my friend Grossmann. Indeed, we had come at that time already very close to the solution that will be given in the following.%
\footnote{\foreignlanguage{ngerman}{``So gelangte ich zu der Forderung einer allgemeineren Kovarianz der Feldgleichungen zur\"uck, von der ich vor drei Jahren, als ich zusammen mit meinem Freunde Grossmann arbeitete, nur mit schwerem Herzen abgegangen war. In der Tat waren wir damals der im nachfolgenden gegebenen L\"osung des Problems bereits ganz nahe gekommen.''}\citenumns[778]{EinsteinA1915Relativitaetstheorie}.}
\end{quote}
Very similarly, he expressed himself in a letter to David Hilbert, written on 18 November 1915, the day of his third memoir, in which he had succeeded in computing the correct value of Mercury's perihelion advance on the basis of field equations $R_{\mu\nu}\propto T_{\mu\nu}$:
\begin{quote}
The difficulty was not in finding generally covariant equations for the $g_{\mu\nu}$; for this is easily achieved with the aid of Riemann's tensor. Rather it was hard to see that these equations are a generalization of Newton's law. This insight I only achieved in the last weeks [...], while I had already considered the only possible generally covariant equations, which now turned out to be the correct ones, already three years ago with my friend Grossmann. Only with a heavy heart did we give them up, since it had appeared to me that their physical discussion had shown their incompatibility with Newton's law.%
\footnote{\foreignlanguage{ngerman}{``Die Schwierigkeit bestand nicht darin allgemein kovariante Gleichungen f\"ur die $g_{\mu\nu}$ zu finden; denn dies gelingt leicht mit Hilfe des Riemann'schen Tensors. Sondern schwer war es, zu erkennen, dass diese Gleichungen eine Verallgemeinerung, und zwar eine einfache und nat\"urliche Verallgemeinerung des Newtonschen Gesetzes bilden. Dies gelang mir erst in den letzten Wochen [...], w\"ahrend ich die einzig m\"oglichen allgemein kovarianten Gleichungen, [die] sich jetzt als die richtigen erweisen, schon vor 3 Jahren mit meinem Freunde Grossmann in Erw\"agung gezogen hatte. Nur schweren Herzen trennten wir uns davon, weil mir die physikalische Diskussion scheinbar ihre Unvereinbarkeit mit Newtons Gesetz ergeben hatte.''}\citenumns[Doc.~148]{CPAE08}.}
\end{quote}
In a letter to his Swiss friend Heinrich Zangger, written on 9 December 1915, just a few days after the final breakthrough, he wrote:
\begin{quote}
The interesting thing is that now the inital hypotheses I made with Grossmann are confirmed, and the most
radical of theoretical requirements materialized. At the time we lacked only a few relations of a formal nature, without which the link between the formulas and already known laws cannot be attained.%
\footnote{\foreignlanguage{ngerman}{``Interessant ist, dass sich nun die ersten Ans\"atze best\"atigen, die ich mit Grossmann machte, und die die radikalsten theoretischen Forderungen realisieren. Es fehlten uns damals nur einige Relationen formaler Art, ohne welche der Anschluss der Formeln an die bereits bekannten Gesetze nicht zu erlangen ist.''}\citenumns[Doc.~Vol.8, 161a]{CPAE10}. Similar statements can be found in letters to Arnold Sommerfeld, 28 November 1915, to Michele Besso, 10 December 1915, and to Hendrik A. Lorentz, 1 January 1916 and 17 January 1916 \citenumns[Docs.~153, 162, 177, 183]{CPAE08}.}
\end{quote}
A few months after the completion of the general theory of relativity by publication, Einstein published a comprehensive exposition of the final theory.\cite{EinsteinA1916Grundlage,SauerT2005Paper} The paper begins with a page-long introductory paragraph, in which Einstein gave credit to the mathematical traditions that he had built upon, singling out the contributions of Minkowski, as well as of Gauss, Riemann, Christoffel, Ricci and Levi-Civita.\footnote{That first page got lost, presumably by accident,\cite{DickensteinA2009Cover} in the preparation of the English translation that was read widely in a Dover edition in the Anglo-Saxon world.} The hommage to the mathematical tradition ends with an expression of gratitude for Grossmann:

\begin{quote}
Finally I want to acknowledge gratefully my friend, the mathematician Grossmann, whose help not only saved me the effort of studying the pertinent mathematical literature, but who also helped me in my search for the field equations of gravitation.%
\footnote{\foreignlanguage{ngerman}{``Endlich sei an dieser Stelle dankbar meines Freundes, des Mathematikers Grossmann, gedacht, der mir durch
seine Hilfe nicht nur das Studium der einschl\"agigen mathematischen Literatur ersparte, sondern mich
auch beim Suchen nach den Feldgleichungen der Gravitation unterst\"utzte.''}%
\citenumns[769]{EinsteinA1916Grundlage}.}
\end{quote}

\section{Biographical Epilogue}

Although their biographical and intellectual trajectories continued largely independently after their collaboration in Zurich, Einstein and Grossmann remained friends. During a visit to the Grossmann family in Zurich in summer 1919, the two friends apparently even talked about the possibility that Einstein would come back to Zurich\citenumns[Vol. 9, 72e, 74d]{CPAE10}. Grossmann tried to lure Einstein back but Einstein was discouraged by the idea of having to face a full teaching load again.

In early January 1920, at the height of the public interest in Einstein's theory of general relativity,
Marcel Grossmann published a two-piece article entitled ``A New Worldview'' in the \emph{Neue Schweizer Zeitung}.\cite{GrossmannM1920Weltbild} Just a few weeks earlier, on 6 November 1919 at a joint session
of the Royal Society and the Royal Astronomical Society in London, it had been announced that the results of the British eclipse expedition had confirmed Einstein's relativistic theory of gravitation. After Grossmann put Einstein's achievement in a line with Galilei, Kopernikus, Kepler, and Newton, he wrote:
\begin{quote}
As a school day friend and fellow student of this great physicist I might be permitted to follow up on the suggestion of the editorial board and give an understanding of the man and the work to a wider audience,
and try to give a concept of the ingenuity and consequences of his ideas. For years already these ideas have engaged his colleagues but it is only in recent months that also in wider intellectual circles it is pointed out that here a revolution was begun and completed of all of our basic concepts in physics, astronomy, geometry, as well as philosophical epistemology.%
\footnote{\foreignlanguage{ngerman}{``Es sei einem Jugendfreund und Studiengenossen dieses gro{\ss}en Physikers gestattet, auf den Wunsch der Redaktion den Menschen und sein Werk auch einem weiteren Leserkreis n\"aher zu bringen, zu versuchen, einen Begriff von der Genialit\"at und Tragweite seiner Ideen zu bieten. Seit Jahren bewegen sie die engeren Fachgenossen, aber erst die letzten Monate haben auch weitere wissenschaftliche Kreise darauf aufmerksam gemacht, da{\ss} hier ein Umsturz aller unserer Grundvorstellungen in Physik, Astronomie, Geometrie und philosophischer Erkenntnistheorie angebahnt und vollendet wurde.''}\cite{GrossmannM1920Weltbild}.}
\end{quote}
Grossmann continued to give some biographical information on Einstein, pointed out that he obtained Swiss citizenship as a student at the ETH and had kept his Swiss nationality even after his move to Berlin. He emphasized Einstein's pacifist stance during the war, mentioning the counter manifesto \citenumns[Doc.~8]{CPAE06} with Georg Friedrich Nicolai (1874--1964) and Wilhelm F\"orster (1832--1921) against the infamous manifesto of the 93. He then gave a short characterization of the special theory of relativity. His discussion of special relativity ends with this observation:
\begin{quote}
One can explain all those premises and consequences of the theory of relativity only to those whose mathematical knowledge and maturity of judgment go far enough. The mathematician possesses in his formal language a shorthand of thinking, which is not only useful but indispensable for more complicated trains of thought. Just as the skill of shorthand writing helps us following a lecture, it is the mathematical language of formulas that enables us to create complicated trains of thought, which could not be brought to a conclusion without it.%
\footnote{\foreignlanguage{ngerman}{``V\"ollig verst\"andlich k\"onnen alle diese Voraussetzungen und Folgerungen aus der Relativit\"atstheorie nur dem gemacht werden, dessen mathematische Kenntnis und Reife des Urteils weit genug gehen. Der Mathematiker besitzt in seiner Formelsprache eine \emph{Stenographie des Denkens}, die ihm n\"utzlich, ja unentbehrlich ist f\"ur kompliziertere Gedankeng\"ange. Wie die Kenntnis der Stenographie das Folgen eines Vortrages erleichtert, so erm\"oglicht die mathematische Formelsprache die Aufstellung verwickelter Gedankenketten, die man ohne sie gar nicht zu Ende denken k\"onnte.''}\cite{GrossmannM1920Weltbild}.}
\end{quote}
With the general theory of relativity, Grossmann wrote, Einstein even topped his own prior achievement. He not only generalized special relativity but also succeeded in drawing experimentally verifiable consequences, which indeed had been confirmed. Einstein's former collaborator, who helped with the mathematics, then observed about the role of mathematics in the genesis of general relativity:
\begin{quote}
Laymen have an entirely misleading conception of the essence of mathematical and generally scientific research. Also in this field of human intellect, something \emph{new} is only being created by \emph{intuition}, by \emph{creative imagination}. The great mathematicians and physicists are not `good calculators,' in this respect they are outplayed by your average able accountant; nor is someone who plays the piano with virtuosity a great musician! Original achievements in all fields of human knowledge and capability are \emph{artistic achievements} and follow their own laws.\\
To a person who witnessed Einstein's first laborious attempts in the years 1912 and 1913, as the composer of these lines did, they must appear like the ascent of an inaccessible mountain in the dark of the night, without path or trail, without foothold or direction. Experience and deduction provided only few and insecure handholds. All the higher we have to value this intellectual deed.%
\footnote{\foreignlanguage{ngerman}{``Der Laie macht sich eben eine ganz falsche Vorstellung davon, was das Wesen mathema\-tischer und allgemein exakt-naturwissenschaftlicher Forschung ist. \emph{Neues} schafft auch auf diesem T\"atigkeitsfeld des menschlichen Geistes nur die \emph{Intuition}, die \emph{sch\"opferische Phantasie}. Die gro{\ss}en Mathematiker und Physiker sind nicht etwa {\glqq}gute Rechner{\grqq}, da werden sie zumeist von jedem t\"uchtigen Buchhalter \"ubertrumpft; das ist ebenso wenig der Fall, als da{\ss} ein gro{\ss}er Musiker sei, wer virtuous Klavier spielen k\"onne! Originelle Leistungen in allen Gebieten des menschlichen Wissens und K\"onnens sind \emph{k\"unstlerische Leistungen} und gehorchen deren Gesetzen.\\
Wer, wie der Schreiber dieser Zeilen, die ersten m\"uhseligen Tastversuche Einsteins in den Jahren 1912 und 1913 miterlebt hat, mu{\ss} an die Eroberung eines schwerzug\"anglichen Berggipfels in dunkler Nacht, ohne Weg und Steg, ohne Halt und ohne Richtung denken. Erfahrung und Dedukion boten nur sp\"arliche und unsichere Griffe. Um so h\"oher ist diese geistige Tat zu werten.''}}
\end{quote}
No claims to co-discovery by Grossmann indeed!

Also in 1920, Grossmann felt compelled to intervene on behalf of the Swiss physicists and mathematicians and to defend Einstein's relativity theory against criticism
by a Bernese physicist. Eduard Guillaume (1881--1959), then a mathematician at the Swiss
Federal Insurance Bureau, had been a colleague of Einstein's at the Swiss
patent office. In 1909, the two had even done some experimental work together \citenumns[Doc.~143]{CPAE05},
and, in 1913, Guillaume had translated into French Einstein's short version of the \emph{Entwurf} presented at
the Frauenfeld meeting.\cite{EinsteinA1914Bases}
But beginning in 1917, Guillaume had started to criticize the special theory of relativity in a number of
articles that appeared mostly in the {\it Archives des sciences physiques et naturelles}.

On 5 February 1920, Grossmann forwarded one of Guillaume's papers at the latter's request \citenumns[Doc.~300]{CPAE09}. Einstein's response was short and harsh: ``Guillaume's notice is stupid
like everything this man dashes off about relativity.''\footnote{``Guillaumes Notiz ist bl\"ode wie alles, was dieser Mann \"uber Relativit\"at zusammen schreibt.'' \citenumns[Doc.~330]{CPAE09}.} Presumably with reference
to this letter, Grossmann had referred to Einstein's opinion about Guillaume in a piece that
he published on 15 June 1920 in the \emph{Neue Schweizer Zeitung}.\cite{GrossmannM1920Physikprofessoren} Guillaume complained about
Grossmann's criticism in private correspondence with Einstein and tried to explain his critical views.
Einstein patiently continued their correspondence but admitted that he was unable to understand
what Guillaume's point really was.\cite{CPAE10,GenovesiA2000Carteggio}

\begin{figure}[h]
\begin{center}
\includegraphics[width=6cm]{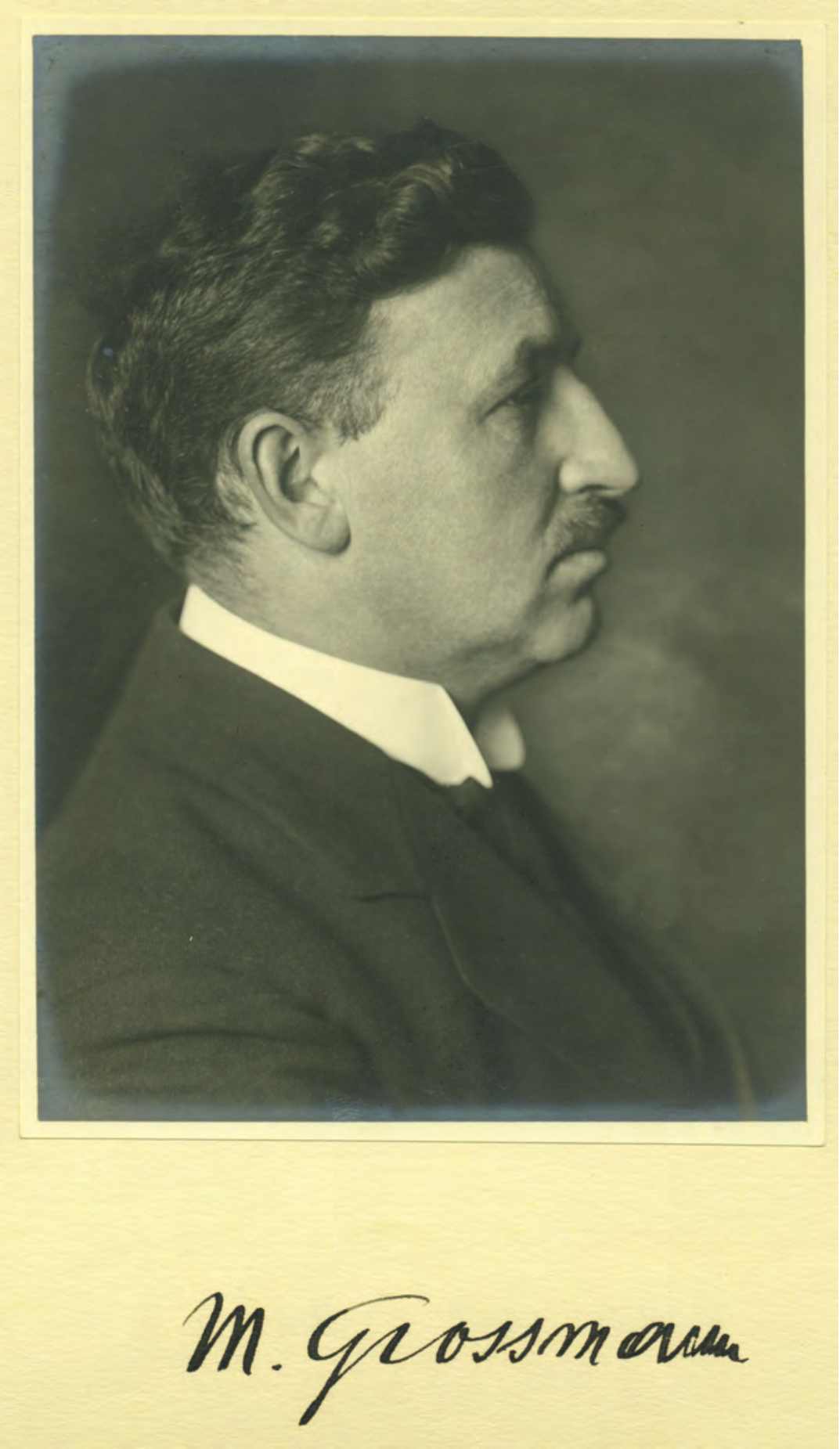}
\caption{Marcel Grossmann (1878--1936). \copyright ETH Bibliothek Z\"urich, Bild\-archiv.}
\label{fig:Grossmann2}
\end{center}
\end{figure}

On 3 September 1920, Grossmann published a note in the \emph{Neue Schweizer Zeitung} accusing Einstein's German colleagues of not supporting him against the anti-Einstein campaign that had just culminated with the infamous lectures by Paul Weyland (1888--1972) and Ernst Gehrcke (1878--1960) in the Berlin Philharmonic.\cite{GrossmannM1920Hetze}
A few days later, on 9 September 1920, Grossmann made another attempt at getting Einstein back to Switzerland: ``Are you still not ripe for Zurich yet?'' he asked in a letter, in which he also pointed out that ``both our boys, who are in the same class at the Gymnasium, are already calculating with logarithms''\citenumns[Doc.142]{CPAE10}. In the letter, he also asked for another statement on Guillaume, which he might then translate into French and forward for publication in Geneva's \emph{Archives des sciences physiques et naturelles}. In Grossmann's words, ``a cult is forming around Guillaume that thinks it must correct essential points of your concepts.'' Einstein was pleased ``that our boys are classmates, like we were,'' and complied with the request by sending a harsh statement on Guillaume's work, stating that he was ``unable to attach any kind of clear sense to Guillaume's explications''\citenumns[Doc. 148]{CPAE10}. Einstein's statement did not get published. Instead, Grossmann himself wrote a brief statement in the \emph{Archives}.\cite{GrossmannM1920Mise} According to his note, Guillaume had developed his interpretation of the Lorentz transformations at the international conference of mathematicians in Strasbourg, and while it was impossible for Grossmann to assess the significance of the theory as far as the physics was concerned, he could put the finger on the mathematical error that he committed.\footnote{``Il m'est impossible de saisir la port\'ee physique de sa th\'eorie; mais du point de vue math\'ematique on peut toucher du doigt l'erreur commise.''\cite{GrossmannM1920Mise}.}
Guillaume claimed that he had found a new invariant of the Lorentz transformations. But one knew that the Lorentz transformations are associated with a line element with constant coefficients, which possesses no invariants at all. Only differential quadratic forms with variable coefficients possess an invariant function, the curvature of space according to Riemann and Christoffel, on which the gravitational field equations are based. Guillaume's expression, on the other hand, was ``an identity pure and simple. It does not carry any physical or mathematical interest.''\footnote{``une identit\'e pure et simple. Elle ne peut pas donc avoir aucun int\'er\^et math\'ematique ou physique.''\cite{GrossmannM1920Mise}.}

For the following years, we have a few items in the Einstein Archives that document Grossmann's and Einstein's ongoing friendship. In July 1922, Einstein jokingly pondered to name a sailing boat that he had bought used and that initially had sunk due to a leakage either ``letdown'' (``Reinfall'')  or ``Grossmann'', presumably with reference to Grossmann's earlier assistance in keeping him afloat, as it were \citenumns[Doc.~306]{CPAE13}. Later that year, Einstein and his (second) wife tried to see Grossmann in Zurich on their way out for their Japan trip but missed him. A postcard that they sent instead also conveys congratulations to Grossmann's son who had just graduated together with Einstein's son from the \emph{Realgymnasium} of the Kantonsschule in Z\"urich.\footnote{With Hans Marcel outperforming Hans Albert by scoring 62 versus 55.5 out of 66 possible points \citenumns[Doc.~382, esp. note 5]{CPAE13}.} In August 1923, Grossmann congratulated Einstein on his decision to rejoin the committee of intellectual cooperation of the League of Nations.\footnote{M.~Grossmann to A.~Einstein, 1 August 1923 (AEA 11 464).} At the end of the year, Grossmann wrote again. This time, he was congratulating Einstein to the good performance of his son who had scored best at examinations at the ETH that he, Grossmann, had taken. He also invited Einstein for a lecture, and he reiterated his offer that Einstein could come back to the ETH if he so wished.\footnote{See A.~Einstein to M.~Grossmann, 28 December 1923 (AEA 11 466).} Another invitation to deliver a lecture at the annual assembly of the \emph{Schweizerische Naturforschende Gesellschaft} to take place in October 1924 in Lucerne was conveyed through correspondence by Grossmann.\footnote{M.~Grossmann to A.~Einstein, 11 January 1924 and 12 March 1924 (AEA 11 469, 11 470).} This time Einstein accepted.\footnote{A.~Einstein to M.~Grossmann, 15 March 1924 (AEA 11 505); in Lucerne Einstein lectured ``on the ether.''\cite{EinsteinA1924Aether}} As mentioned above, Einstein visited Grossmann again in summer 1925, as, indeed, he might have done more frequently than is documented, whenever he visited Zurich to see his sons.

Grossmann's last scientific publication concerns a mathematical critique of the geometric foundation of Einstein's so-called teleparallel approach to a unified field theory of gravitation and electromagnetism.\cite{GrossmannM1931Fernparallelismus} In summer 1928, Einstein had published two short notes in the Proceedings of the Prussian Academy, in which he introduced his new approach and its underlying geometric idea.\cite{EinsteinA1928Riemann,EinsteinA1928Moeglichkeit,SauerT2006Equations} The geometry of teleparallelism was, in fact, not new at all, but had been investigated by others before, notably by \'Elie Cartan (1869--1951) in the early twenties. It was formulated then in terms of tetrad fields, orthogonal vector fields $h_{a\mu}$ defined at each point of a manifold such that
\begin{equation}
\sum_a h_{a\mu}h^{a\nu} = {\delta_{\mu}}^{\nu}; \qquad \sum_{\mu}h_{a\mu}{h_{b}}^{\mu} = \delta_{ab}
\end{equation}
and a given metric $g_{\mu\nu}$ is expressed as
\begin{equation}
g_{\mu\nu} = \sum_s{h^s}_{\mu}h_{s\nu}.
\end{equation}
Here Latin letters denote the tetrad indices and Greek indices denote the coordinate indices.
Einstein's point in that approach was that the tetrads both allow a geometrical interpretation of distant parallelism on the manifold but also carry more degrees of freedom than the metric tensor field. Specifically,
the tetrad fields allow the definition of an antisymmetric connection
\begin{equation}
\Pi^{\nu}_{\mu\sigma} = \sum_ah^{a\nu}h_{a\mu,\sigma},
\label{eq:metricintermsoftetrads}
\end{equation}
(later called the Weitzenb\"ock connection)
which gives rise to a flat Ricci tensor
\begin{equation}
{P^{\iota}}_{\kappa\lambda\mu} = - \Pi^{\iota}_{\kappa\lambda,\mu} + \Pi^{\iota}_{\kappa\mu,\lambda}
+ \Pi^{\iota}_{\alpha\lambda}\Pi^{\alpha}_{\kappa\mu} - \Pi^{\iota}_{\alpha\lambda}\Pi^{\alpha}_{\kappa\mu}\equiv 0
\end{equation}
(where comma-separated indices denote coordinate derivatives and the summation convention applies).
Einstein was hoping to exploit the additional degrees of freedom provided by the tetrad fields, and the new geometric interpretation of distant parallelism, for a unified description of the gravitational and electromagnetic fields.\cite{SauerT2006Equations} He published a number of further papers on the theory in the following two years, and in early 1929, the new approach also made headlines in the daily press, for reasons not altogether rational but indicative of Einstein's celebrity fame in Weimar Germany. In 1930, Einstein spoke about the approach at the so-called \emph{Weltkraft-Konferenz}, which took place in Berlin from 16 to 25 June 1930.\cite{EinsteinA1930Problem}

It might have been the high visibility of Einstein's new theory in the public, which prompted Grossmann to take an interest in it. In any case, he asked Einstein for offprints of the theory, which Einstein sent him with an accompanying letter on 14 August 1930 (AEA 11 507). After studying Einstein's papers, Grossmann wrote back that the mathematical foundations on which Einstein intended to erect his ``grandiose edifice'' was in his opinion ``an illusion.'' He doubted the very existence of a ``pseudo-Euclidean'' manifold. Either the metric had constant coefficients, in which case the Riemann curvature would vanish and the manifold would be Euclidean, or the Riemann curvature was non-vanishing, in which case the manifold could not be flat (AEA 11 472).

Einstein responded promptly with a letter justifying his approach (AEA 11 509). He illustrated the concepts by considering a general curved two-dimensional surface embedded in three-dimensional space. He pointed out that the usual Christoffel symbols followed from the law of parallel transport, if the latter was a) metric preserving and if b) the connection was assumed to be symmetric. In his approach, however, the connection was not assumed to be symmetric, instead he demanded integrability. Grossmann was not convinced, and on 13 September 1930, Einstein sent him another letter, explaining the different concepts of ``parallel'' in his theory (AEA 11 476).

Again, Grossmann was not convinced. In a short letter of 23 November 1930, he argued that in Einstein's case, the metric $g_{\mu\nu}$ would necessarily be asymmetric, violating a basic assumption of Riemannian geometry (AEA 11 475). In that letter, Grossmann also announced a publication of his own, and asked Einstein whether he preferred to publish it in the Proceedings of the Berlin Academy or whether he should publish it in Switzerland.

Grossmann's last publication was dated 16 January 1931 and published in the quarterly journal of the \emph{Naturforschende Gesellschaft} in Zurich. It was entitled ``Distant parallelism? Correction of the chosen foundation for a unified field theory.''\footnote{``Fernparallelismus? Richtigstellung der gew\"ahlten Grundlage f\"ur eine einheitliche Feldtheorie.''\cite{GrossmannM1931Fernparallelismus}.} The paper begins like this:
\begin{quote}
My dear friend Albert Einstein, member of the Prussian Academy, has been striving for some time to lay the foundations for a unified field theory of gravitation and electromagnetism in the framework of the general theory of relativity. This aim is formidable and worthy to be pursued.%
\footnote{\foreignlanguage{ngerman}{``Mein lieber Freund Albert Einstein, Mitglied der preussischen Akademie der Wissenschaften, bem\"uht sich seit einiger Zeit im Rahmen der allgemeinen Relativit\"atstheorie eine \emph{einheitliche Feldtheorie} f\"ur Gravitation und Elektromagnetismus zu begr\"unden. Dieses Ziel ist hochgesteckt und wohl wert, angestrebt zu werden.''}\citenumns[42]{GrossmannM1931Fernparallelismus}.}
\end{quote}
He agreed that such a foundation would give theoretical physics logical necessity, unity, and consistency, and he admired Einstein for his energy, fantasy, and persistence in pursuing this aim. But health impediments had kept him apart from scientific life, and only shortly ago had he heard about Einstein's lecture at the \emph{Weltkraft-Konferenz} and gotten hold of its printed version. Reading Einstein's lecture he then observed:

\begin{quote}
At the end of these profound demonstrations, Einstein touched in passing on the notion of ``distant parallelism'' and this gave me \emph{immediate} pause. A perusal of the main physical and mathematical works led me to a \emph{complete rejection} of this and other concepts.%
\footnote{\foreignlanguage{ngerman}{``Am Schlusse dieser tiefsch\"urfenden Ausf\"uhrungen wurde auch der
Begriff des {\glqq}Fernparallelismus{\grqq} gestreift und machte mich \emph{sofort} stutzen. Eine Durchsicht der haupts\"achlichen physikalischen und mathematischen Abhandlungen f\"uhrte mich zur \emph{v\"olligen Ablehnung} dieses und anderer Begriffe.''}\citenumns[43]{GrossmannM1931Fernparallelismus}.}
\end{quote}
In the body of the paper, Grossmann argues against the logical soundness and consistency of Levi-Civita's concept of ``parallelismo assoluto'' of 1917, of Cartan's ``parall\'elisme absolu'' of 1922, and of the concepts of ``Fernparallelismus'' of 1928 and its later invariant-theoretic characterization. Grossmann emphasized that he was not arguing against relativity theory as such, nor against the program of finding a unified field theory but only against the geometric structure that was given to it. Grossmann's main argument in his paper was based on taking Felix Klein's (1849--1925) so-called ``Erlangen Program'' as a criterion for any acceptable geometry. Specifically, he argued that the Erlangen program demanded that group theoretic invariance of infinitesimal transformations of the manifold as a criterion for the characterization a given geometry would need to hold \emph{globally} for the entire manifold. A detailed analysis of Grossmann's critique would be beyond the scope of the present article, in the following I will only discuss this paper with a view toward the relationship between Grossmann and Einstein.

Grossmann began his critical comments of Einstein's teleparallelism with a reference to their earlier correspondence:
\begin{quote}
This fall (1930), I had an opportunity to raise my doubts vis-a-vis Einstein. With his old friendship and
loyalty he took pains repeatedly to try to explain to me his point of view. But both of us dug in our heels---an oddity, considering that this concerns a purely mathematical matter of dispute, only to be explained by my physical problems to make myself clear.%
\footnote{\foreignlanguage{ngerman}{``Ich hatte diesen Herbst (1930) Gelegenheit, ihm meine Zweifel zu \"aussern. In alter Freundschaft und
Anh\"anglichkeit gab er sich wiederholt M\"uhe, mich zu bekehren. aber wir beide verharrten hartn\"ackig auf unserm Standpunkt,---eine Merkw\"urdigkeit bei einer im wesentlichen mathematischen Streitfrage und nur zu erkl\"aren aus der k\"orperlichen M\"uhe meinerseits, mich verst\"andlich zu machen.''}\citenumns[50]{GrossmannM1931Fernparallelismus}.}
\end{quote}
Grossmann's main point of contention was to insist on the theorem that a manifold is flat (``Euclidean'') if its curvature tensor vanishes, and the latter is the case, if the coefficients of the metric happen to be all constants. Grossmann did not accept the concept of a Ricci flat manifold that was called ``non-Euclidean'' because it carried non-vanishing torsion. His criticism and confusion may have had a point, perhaps, in that the distinction between two different connections on the same manifold, one torsion-free with non-vanishing Riemann curvature, the other Riemann flat with non-vanishing torsion may not yet have been entirely understood.

In any case, it appears that Grossmann kept thinking in terms of standard Riemannian geometry and conceived of Einstein's so-called ``pseudo-Euclidean'' manifolds in analogy to flat two-dimensional surfaces, carrying extrinsic curvature embedded in three-dimensional space. He wrote:
\begin{quote}
Einstein rejoices over the simplicity of his ``pseudo-Euclidean'' geometry. But also this circumstance is inconclusive, since it only expresses the logical consistency of Euclidean geometry, which appears in an invariant-theoretically general way, i.e., it comprises all bendings of the Euclidean plane and therefore is not fully recognized.
We have seen Einstein before---it was in the year 1913---publish `field equations' following this method, which had to be modified a few years later; at that time, I was also responsible.%
\footnote{\foreignlanguage{ngerman}{``Einstein freut sich \"uber die Einfachheit seiner
{\glqq}pseudo-euklidischen{\grqq} Geometrie. Auch dieser Umstand hat keine Beweiskraft, denn es kommt in ihm eben die \emph{Widerspruchslosigkeit der euklidischen Geometrie} zum Ausdruck, die in invarianten-theoretisch allgemeiner Form erscheint, also alle Verbiegungen der euklidischen Ebene mitumfasst und daher nicht erkannt wird.\\
Schon einmal hat Einstein---es war im Jahre 1913---nach dieser Methode {\glqq}Feldgleichungen{\grqq} ver\"offentlicht, die nach wenigen Jahren abge\"andert werden mussten; damals war ich mitverantwortlich.''}\citenumns[54]{GrossmannM1931Fernparallelismus}.}
\end{quote}
Toward the end of his paper, Grossmann wrote:
\begin{quote}
As students, we, Albert Einstein and I, often analysed psychologically joint acquaintances as well as ourselves. During one of those conversations he once made the accurate observation: your main weakness is, you cannot say `no.' Well, in the meantime, I learned to say `no' and did so profoundly and frequently, not always to the joy and satisfaction of my fellow human beings. There exist in science, in the educational system, in politics, and in life generally phenomena about which you can only shake your head, even if you see their true causes. Also to the development of differential geometry and mathematical physics over the last years, I am saying here `no,' because I am convinced that this is in the interest of science and utlimately also in the interest of my friend.%
\footnote{\foreignlanguage{ngerman}{``Als Studenten haben wir, Albert Einstein und ich, oft gemeinsame Bekannte und einander gegenseitig psychologisch zergliedert. Bei einem solchen Gespr\"ach machte er mir einmal die treffende Bemerkung: Dein Hauptfehler ist, du kannst nicht {\glqq}nein{\grqq} sagen. Nun, in der Zwischenzeit habe ich es erlernt und zwar gr\"undlich und oft getan, zumeist nicht nur zur Freude und Genugtuung meiner Mitmenschen. Es gibt eben in der Wissenschaft, im Unterrichtswesen, in der Politik und im Leben \"uberhaupt Erscheinungen, zu denen man nur den Kopf sch\"utteln kann, auch wenn man sie in ihrer wahren Ursache durchschaut. Auch zur Entwicklung, welche die Differentialgeometrie und die mathematische Physik seit einigen Jahren genommen haben, sage ich hier {\glqq}nein{grqq}, aus der \"Uberzeugung heraus, damit der Wissenschaft und, ends aller Enden, auch meinem Freunde Vorschub zu leisten.''}\citenumns[58--59]{GrossmannM1931Fernparallelismus}.}
\end{quote}
To end his paper on a positive note, Grossmann suggested a mathematical research that would reconsider the foundations of Riemannian geometric manifolds.

Grossmann's final paper is an odd contribution that clearly reflects the pains of his illness. Despite its stylistic oddity and sweeping criticism of major advancements in differential geometry, it nevertheless might perhaps be seen to advance a valid point. Grossmann rightly points out that the geometric foundations of torsion geometry violate standard Riemannian geometry, and he calls for a systematic reflection. The undifferentiated co-existence of two concepts of connections in Einstein's teleparallelism called for such clarification, even if Grossmann did not clearly see what went wrong.  It should also be pointed out that Grossmann's critical comments contain a rather basic blunder. He quoted from Einstein's earlier correspondence and explicitly observed that Eq.~(\ref{eq:metricintermsoftetrads}) expressing the metric in terms of the tetrads would entail that the metric be asymmetric, thus undermining the very foundations of Riemannian geometry. As was pointed out to Grossmann by Einstein and others, that observation is, of course, simply wrong, and Grossmann later admitted his mistake (AEA 11 480).

When Grossmann sent an offprint of his published paper to Einstein in April 1931 (AEA 11 478), Einstein replied a few weeks later with a long and friendly letter, in which he refused to enter into a controversy in print (AEA 11 513). Instead, he took pains, again, to explain to Grossmann the rationale of his new approach and much more clearly distinguished between the Levi-Civita connection and the Weitzenb\"ock connection. It is the last direct item of correspondence between the two friends extant in the Einstein Archives.

When word of Grossmann's passing reached him a few years later in Princeton, Einstein sent a condolence letter to his widow (AEA 70 394, \citenumns[177]{SeeligC1952Einstein}). In warm words, he expressed his appreciation and admiration for her loyalty to her husband and for her sacrifices in looking after him, and then he reminisced about Grossmann:
\begin{quote}
The joint years as students come back to mind---he a masterful student, I disorderly and dreamy. He connected with the professors and grasped everything easily, I was aloof and dissatisfied, not much liked. But we were good friends, and the conversations over iced coffee in the \emph{Metropol} every few weeks are among my fondest memories. Then end of studies---I was alone of a sudden, facing life helplessly. But he stood by me and through him (and his father) I came to Haller at the patent office a few years later. It was a kind of livesaving, without which I might not have died but certainly would have withered intellectually. A decade later, joint, feverish scientific work on the formalism of the general theory of relativity. It was not completed, since I moved to Berlin, where I continued work on my own. Then soon came his illness, early signs showing already during the studies of my son Albert. Often and with pain I thought of him, but we saw each other only rarely when I was visiting him.%
\footnote{``Da steigt die gemeinsame Studentenzeit herauf---er meisterhafter Student, ich unordentlich und vertr\"aumt. Er verbunden mit den Lehrern
und alles leicht fassend, ich abseits und unbefriedigt, wenig beliebt. Aber wir waren gute Freunde, und die Gespr\"ache beim Eiskaffee
im Metropol alle paar Wochen geh\"oren zu meinen h\"ubschesten Erinnerungen. Dann Ende der Studien---ich pl\"otzlich von allen verlassen, ratlos vor dem Leben stehend. Er aber stand zu mir und durch ihn (und seinen Vater) kam ich ein paar Jahre sp\"ater zu Haller ans Patentamt. Es war eine Art Lebensrettung, ohne die ich wohl zwar nicht gestorben aber geistig verk\"ummert w\"are.\\
Ein Jahrzehnt sp\"ater die gemeinsame fieberhafte wissenschaftliche Arbeit um den Formalismus der allgemeinen Relativit\"at. Es blieb unvollendet, weil ich nach Berlin ging, wo ich nun allein weiterarbeitete. Dann kam bald seine Krankheit, w\"ahrend meines Sohnes Alberts
Studium zeigten sich schon die Vorboten. Viel und schmerzlich gedachte ich seiner, aber wir sahen uns nur mehr selten, wenn ich zu Besuch dort war.''\citenumns[177]{SeeligC1952Einstein}.}
\end{quote}
Einstein then added some comments on his own age and experience and then concluded his condolences by saying:
\begin{quote}
But one thing is beautiful. We were friends and remained friends throughout life.%
\footnote{``Aber eines ist doch sch\"on. Wir waren und blieben Freunde durchs Leben hindurch.'' ibid.}
\end{quote}


\section*{Acknowledgments}

I wish to thank Matthias Blau, Christian Bracco, Marc Linder, David Rowe, John Stachel, and Andreas Verdun for helpful comments on earlier versions of this paper. Special thanks to Tim R\"az for many interesting discussions, and to Robert Jantzen for his encouragement and stylistic advice.


\bibliographystyle{unsrt}
\bibliography{mg13}

\end{document}